%% file: GLEAM-X.tex
\documentclass{pasa}%

\usepackage{graphicx}
\usepackage{xcolor}
\usepackage{amsmath}	% Advanced maths commands
\usepackage{amssymb}	% Extra maths symbols
\usepackage{rotating}
\usepackage{supertabular,booktabs}

\newcommand{\myaffil}[1]{$^{\rm #1}$}
\newcounter{inst}
\newcommand{\inst}[1]{\noindent%
   \refstepcounter{inst}\myaffil{\arabic{inst}\label{#1}}     % comment if you prefer letter references
   }
   
\newcommand{\arcdeg}{\ensuremath{^{\circ}}}

\newcommand{\perbeam}{\,beam\ensuremath{^{-1}}}
\DeclareMathOperator{\arcsinh}{arcsinh}

\newcommand{\ncol}{722}
\newcommand{\survarea}{1,447}
\newcommand{\nsrc}{78,967}

\newcommand{\nfit}{71,320}
\newcommand{\nplfit}{70,432}
\newcommand{\ncplfit}{888}
\newcommand{\srcdensity}{55}
\newcommand{\pctreliablelow}{98.2}
\newcommand{\pctreliablehigh}{99.7}

\newcommand{\fig}{Fig.}

\newcommand{\Fig}{Fig.}
\newcommand{\Figs}{Figs.}
\newcommand{\sect}{Section}

\newcommand{\Sect}{Section}
\newcommand{\Section}{Section}

\newcommand{\Tab}{Table}

\newcommand{\Eqn}{Equation}
\newcommand{\Eqns}{Equations}

\newcommand{\farcm}{\mbox{\ensuremath{.\mkern-4mu^\prime}}}%    % fractional arcminute symbol: 0.'0
\newcommand{\arcsec}{\mbox{\ensuremath{''}}} % arcsecond symbol

\title[GLEAM-X Survey Description]{GaLactic and Extragalactic All-sky Murchison Widefield Array survey eXtended (GLEAM-X) I: Survey Description and Initial Data Release}

\author[Hurley-Walker~et~al.]{N.~Hurley-Walker\myaffil{\ref{ICRAR}},
T.~J.~Galvin\myaffil{\ref{ICRAR},\ref{CASS}},
S.~W.~Duchesne\myaffil{\ref{ICRAR},\ref{CASS}},
X.~Zhang\myaffil{\ref{CASS},\ref{SHAO}},
J.~Morgan\myaffil{\ref{ICRAR}},
P.~J.~Hancock\myaffil{\ref{ICRAR},\ref{CIC}},
T.~An\myaffil{\ref{SHAO}},
T.~M.~O.~Franzen\myaffil{\ref{ASTRON}},
G.~Heald\myaffil{\ref{CASS}},
K.~Ross\myaffil{\ref{ICRAR}},
T.~Vernstrom\myaffil{\ref{CASS},\ref{ICRARUWA}},
G.~E.~Anderson\myaffil{\ref{ICRAR}},
B.~M.~Gaensler\myaffil{\ref{Toronto}},
M.~Johnston-Hollitt\myaffil{\ref{CIC}},
D.~L.~Kaplan\myaffil{\ref{UWisc}},
C.~J.~Riseley\myaffil{\ref{CASS},\ref{Bologna},\ref{INAF}},
S.~J.~Tingay\myaffil{\ref{ICRAR}},
M.~Walker\myaffil{\ref{ICRAR}}\\
{\small \myaffil{}\,Email: nhw@icrar.org}\\
{\small \inst{ICRAR}\,International Centre for Radio Astronomy Research, Curtin University, Bentley, WA 6102, Australia}\\
{\small \inst{CASS}\,CSIRO Space \& Astronomy, PO Box 1130, Bentley WA 6102, Australia}\\
{\small \inst{SHAO}\,Shanghai Astronomical Observatory, Chinese Academy of Sciences, 80 Nandan Rd, Shanghai, 200030, China}\\
{\small \inst{CIC}\,Curtin Institute for Computation, Curtin University, GPO Box U1987, Perth WA 6845}\\
{\small \inst{ASTRON}\,ASTRON, Netherlands Institute for Radio Astronomy, Oude Hoogeveensedijk 4, 7991 PD, Dwingeloo, The Netherlands}\\
{\small \inst{ICRARUWA}\,International Centre for Radio Astronomy Research, The University of Western Australia, 35 Stirling Hwy, 6009 Crawley, Australia}\\
{\small \inst{Toronto}\,Dunlap Institute for Astronomy and Astrophysics, 50 St. George St, University of Toronto, ON M5S 3H4, Canada}\\
{\small \inst{UWisc}\,Department of Physics, University of Wisconsin--Milwaukee,
Milwaukee, WI 53201, USA}\\
{\small \inst{Bologna}\,Dipartimento di Fisica e Astronomia, Universit\`a degli Studi di Bologna, via P. Gobetti 93/2, 40129 Bologna, Italy}\\
{\small \inst{INAF}\,INAF -- Istituto di Radioastronomia, via P. Gobetti 101, 40129 Bologna, Italy}\\
}

\jid{PASA}
\doi{10.1017/pas.\the\year.xxx}
\jyear{\the\year}

\usepackage{aas_macros}
\usepackage{hyperref} 
\hypersetup{colorlinks,citecolor=blue,linkcolor=blue,urlcolor=blue}

\begin{document}

\begin{frontmatter}
\maketitle

\begin{abstract}
We describe a new low-frequency wideband radio survey of the southern sky. Observations covering 72--231\,MHz and Declinations south of $+30^\circ$ have been performed with the Murchison Widefield Array ``extended'' Phase~\textsc{II} configuration over 2018--2020 and will be processed to form data products including continuum and polarisation images and mosaics, multi-frequency catalogues, transient search data, and ionospheric measurements. From a pilot field described in this work, we publish an initial data release covering \survarea{}\,deg$^2$ over 4\,h$\leq$ RA$\leq 13$\,h, $-32.7^\circ \leq \mathrm{Dec} \leq -20.7^\circ$. We process twenty frequency bands sampling 72--231\,MHz, with a resolution of $2'$--$45''$, and produce a wideband source-finding image across 170--231\,MHz with a root-mean-square noise of $1.27\pm0.15$\,mJy\,beam$^{-1}$. Source-finding yields \nsrc{}~components, of which \nfit{} are fitted spectrally.
The catalogue has a completeness of 98\,\% at $\sim50$\,mJy, and a reliability of \pctreliablelow{}\,\% at $5\sigma$ rising to \pctreliablehigh{}\,\% at $7\sigma$.
A catalogue is available from Vizier; images are made available on the GLEAM-X VO server and SkyView. This is the first in a series of data releases from the GLEAM-X survey.
\end{abstract}


\end{frontmatter}

%%%%%%%%%%%%%%%%%%%%%%%%%%%%%%%%%%%%%%%%%%%%%%%%%%

%%%%%%%%%%%%%%%%% BODY OF PAPER %%%%%%%%%%%%%%%%%%

\section{Introduction}
\begin{figure}
    \centering
    \includegraphics[width=1\linewidth]{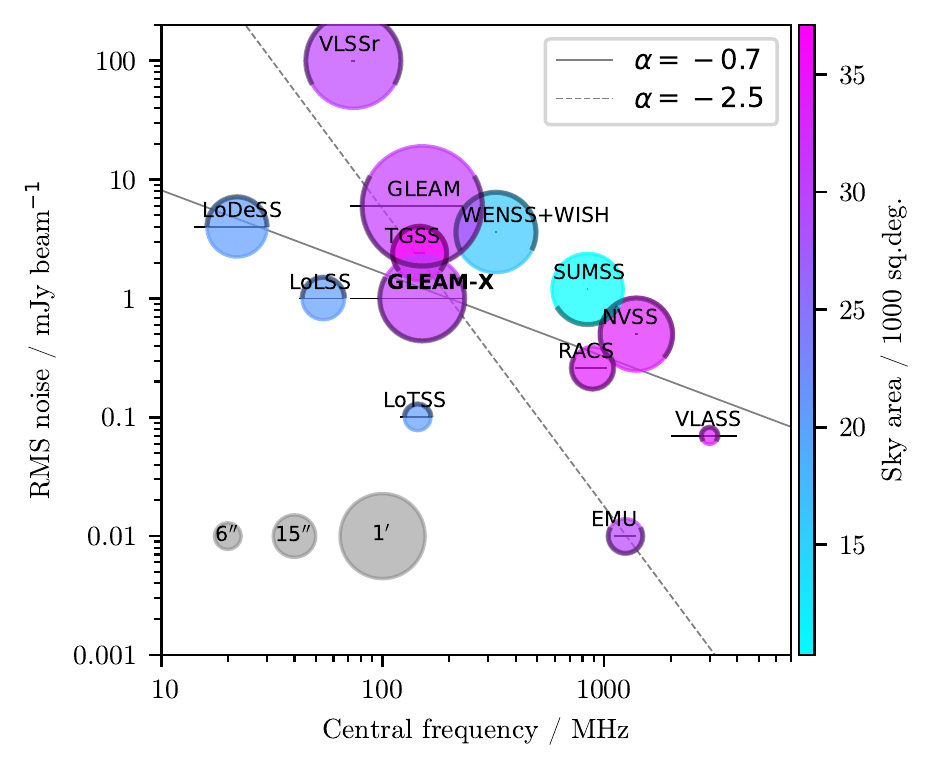}
    \caption{Summary of the sensitivity, frequency, resolution, and sky coverage of a selection of recent and planned large-area radio surveys. The size of the markers is proportional to the survey resolution (full-width-half-maximum of the restoring beam; examples shown in the lower left corner) and their colors show the sky coverage planned. The darkened edges of each marker show the Declination coverage of each survey. The width of each horizontal line shows the frequency range covered by that survey. Representative solid ($\alpha=-0.7$) and dashed ($\alpha=-2.5$) lines show the expected brightness at different frequencies for sources of brightness 1\,mJy\,beam$^{-1}$ at 200\,MHz.
    }
    \label{fig:surveys}
\end{figure}

Radio sky surveys offer a view of the high-energy sky, probing synchrotron, cyclotron, and thermal processes across a range of distances, from planets and exoplanets to high-redshift radio galaxies. At lower frequencies, the fields of view of radio telescopes are larger, enabling large-scale surveys of the radio sky, such as the National Radio Astronomy Observatory (NRAO) Very Large Array (VLA) Sky Survey \citep[NVSS;][]{1998AJ....115.1693C} at 1.4\,GHz, the Sydney University Molonglo Sky Survey \citep[SUMSS;][]{1999AJ....117.1578B,2003MNRAS.342.1117M}, and the Low-frequency Sky Survey Redux at 74\,MHz \citep[VLSSr;][]{2014MNRAS.440..327L}. Spurred by the development of the Square Kilometre Array (SKA), new radio telescopes are exploring the radio sky across wider areas and frequency ranges than accessible in the past (\Fig~\ref{fig:surveys}). 

The Murchison Widefield Array \citep[MWA;][]{2013PASA...30....7T}, operational since 2013, is a precursor to the low-frequency component of the SKA, which will be the world's most powerful radio telescope. The GaLactic and Extragalactic All-sky MWA \citep[GLEAM;][]{2015PASA...32...25W} survey observed the whole sky south of declination (Dec) $+30^\circ$ from 2013 to 2015 between 72 and 231\,MHz. GLEAM has been processed in a multitude of ways: continuum data releases cover most of the extragalactic sky \citep[GLEAM ExGal;][]{2017MNRAS.464.1146H}, the Magellanic Clouds \citep{2018MNRAS.480.2743F}, the Galactic Plane \citep[GLEAM GP;][]{2019PASA...36...47H}, and a deep region over the South Galactic Pole \citep[GLEAM SGP;][]{2021PASA...38...14F}; and polarisation products include all-sky circular \citep{2018MNRAS.478.2835L} and linear polarisation surveys \citep[Polarised GLEAM Survey (POGS);][]{2018PASA...35...43R,2020PASA...37...29R}. Cross-identifications have been provided for the 1,863 brightest radio sources in the mid-infrared \citep[the G4Jy Sample][]{2020PASA...37...17W,2020PASA...37...18W}, and for 1,590 galaxies in the 6dF Galaxy Survey \citep{2021PASA...38...41F}.

While GLEAM had lower sensitivity and resolution than other surveys of the time \citep[e.g. the First Alternative Data Release of the Tata Institute for Fundamental Research Giant Metrewave Radio Telescope Sky Survey:  TGSS-ADR1;][]{2017A&A...598A..78I}, its major advancement was in leveraging its low frequency and very large fractional bandwidth. Extremely steep spectral indices ($\alpha<-2$, for $S\propto\nu^\alpha$) indicate old emission, such as that found in the remnant stage of radio galaxy life cycles \citep{2015MNRAS.447.2468H,2019PASA...36...16D} or ``fossil'' emission in galaxy clusters \citep{2020ApJ...891....1G}; rising spectral indices point toward thermal emission such as found in planetary nebulae \citep{2019PASA...36...47H}. In this frequency range, absorption effects become important for many sources, allowing measurements to probe synchrotron and free-free absorption in extragalactic radio sources \citep{2017ApJ...836..174C} and in Galactic \textsc{Hii} regions \citep{2017MNRAS.465.3163S}. Additionally, GLEAM's very high sensitivity to large angular scales, often resolved out by interferometric surveys, enabled exploration of diffuse emission such as Galactic supernova remnants \citep[e.g.][]{2019PASA...36...48H} and in clusters of galaxies \citep[e.g.][]{2018MNRAS.479..730Z}.

In 2017 the MWA underwent an upgrade to ``Phase~\textsc{II}'', in which an additional 128~tiles were added to the observatory \citep{2018PASA...35...33W}. This enabled observing using two different 128-tile configurations: ``compact'', comprising many redundant baselines to improve calibration toward statistical detection of the Epoch of Reionisation \citep{2018AJ....156..285J}, and ``extended'', an array optimised for imaging (within the constraints of the observatory) with maximum baselines of 5.5\,km, approximately doubling the resolution of the telescope. The latter layout considerably reduces the sidelobes of the synthesised beam, allowing a more ``natural'' weighting of the visibility data, which thereby improves the sensitivity of the instrument; sidelobe confusion is also reduced. The smaller main lobe of the synthesised beam reduces the classical confusion limit from $\sim2$\,mJy to $\sim0.3$\,mJy at 200\,MHz \citep{2019PASA...36....4F}. These improvements make it more feasible to integrate for longer times and thereby reach lower noise levels without quickly approaching a confusion floor.

While GLEAM enabled a huge range of science outcomes, better modeling of the foregrounds for searches for the Epoch of Reonisation, and flux density scale calibration of the low-frequency southern sky, it is fundamentally limited by its low ($\sim2'$) resolution and the sensitivity limits of the original configuration of the MWA. We therefore undertook a wide-area survey with the Phase~\textsc{II} extended array to create GLEAM-X, a deeper, higher-resolution successor to GLEAM, with the same sky and frequency coverage, observed over 2018--2020. During that time, the Long Baseline Epoch of Reionisation Survey \citep[LoBES;][]{2021PASA...38...57L} has demonstrated the survey capability of Phase~\textsc{II} by measuring the spectral behaviour of 80,824~sources over 100--230\,MHz in 3,069\,deg$^2$, down to a noise limit of $\sim2$\,mJy\,beam$^{-1}$, showing the utility of wide-area surveys with the extended array. New radio southern-sky surveys across 800--1400\,MHz using the Australian SKA Pathfinder \citep[ASKAP;][]{2021PASA...38....9H} such as the Rapid ASKAP Continuum survey \citep[RACS;][]{2020PASA...37...48M,2021PASA...38...58H} have also been developed, offering improved morphological information for millions of radio sources.

\Fig~\ref{fig:surveys} shows that the sensitivity of GLEAM-X to ordinary radio galaxies ($-0.8 \lesssim \alpha \lesssim -0.5$) is competitive with other ongoing wide-area surveys such as RACS and the Very Large Array Sky Survey at 3\,GHz \citep[VLASS;][]{2020PASP..132c5001L}. Note also that its sensitivity to steep-spectrum sources ($\alpha=-2.5$) is the same as the upcoming Evolutionary Map of the Universe, which will approach the confusion brightness limit at its frequency \citep[EMU;][]{2011PASA...28..215N,2021PASA...38...46N}. Covering the northern sky at $6$--$60''$ resolution, the LOw Frequency ARray \citep[LOFAR;][]{2013A&A...556A...2V} is observing several ongoing surveys: the LOFAR Two-metre Sky Survey \citep[LoTSS;][]{2017A&A...598A.104S}, the LOFAR Low-Band Array Sky Survey \citep[LoLSS;][]{2021A&A...648A.104D}, and the LOFAR Decametre Sky Survey (LoDeSS; van Weeren et al. in prep).

To reach noise levels that are a significant improvement over GLEAM while still covering a wide area, we accumulate a large ($\sim2$\,PB) volume of visibility data. Releasing processed data products in stages will be of more use to the community than a single data release in the future. This paper is therefore the first in a series of data releases.  We release here a pilot survey area that indicates the qualities that can eventually be expected over the full survey, covering \survarea{}\,deg$^2$ over 4\,h$\leq$ RA$\leq 13$\,h, $-32.7^\circ \leq \mathrm{Dec} \leq -20.7^\circ$. Polarisation processing and an associated early data release will be described in a companion paper, Zhang et al. (in prep). Herein we describe the GLEAM-X observations (\Sect~\ref{sec:obs}), processing pipeline to produce images and mosaics (\Sect~\ref{sec:dr}), source-finding to generate catalogues (\Sect~\ref{sec:catalogue}), and motivate several extensions to the pipeline (\Sect~\ref{sec:extensions}). \Sect~\ref{sec:conclusions} concludes with an outlook on scientific advances enabled by the survey, and plans for further data releases.

%Conventions used throughout this paper are as follows:
All positions given in this paper are in J2000 equatorial coordinates.

\section{Observations}\label{sec:obs}

GLEAM used a drift scan survey strategy to quickly and efficiently observe the entire sky south of Dec~$+30^\circ$ using the Phase~\textsc{I} ``128T'' configuration of the MWA \citep{2015PASA...32...25W}. In the first year (2013-08 -- 2014-06) observations were made along the meridian (HA$=0$\,h), using seven pointings at Declinations centred on $-72^\circ$ to $18.6^\circ$. In the second year, further observations at HA$=\pm1$\,h were taken. By combining the GLEAM data in the image plane over the full range of HA for a region around the South Galactic Pole, \cite{2021PASA...38...14F} were able to reach a noise level of 5\,mJy\,\perbeam{} at 215\,MHz, about half that of the extragalactic data release by \cite{2017MNRAS.464.1146H}, showing that such a strategy was effective.

GLEAM-X therefore adopted a similar strategy, iterating through the same Declination and HAs as GLEAM, but doubling the number of HA$=0$\,h observations, and using the extended configuration of the Phase~\textsc{II} MWA. Observations were performed in month-long blocks in order to observe similar ranges in RA across the different Declination and HAs, making it easier to combine many drift scans in large mosaics in simple sky projections, improving the uniformity of sensitivity across the sky.

To cover 72--231\,MHz using the 30.72-MHz instantaneous bandwidth of the MWA, five frequency ranges of 72--103\,MHz, 103--134\,MHz, 139--170\,MHz, 170--200\,MHz, and 200--231\,MHz were cycled through sequentially, changing every two minutes. Gain calibrators were visited on an hourly basis in order to provide a back-up in case of unsuccessful in-field calibration (\Sect~\ref{sec:calibration}).

After the first observing run in the 2018-A observing semester\footnote{\url{https://www.mwatelescope.org/data/observing}}, the data were triaged to search for poor ionospheric conditions that would hinder high-quality imaging. We determined calibration solutions for the gain calibrator observations on 30-s cadences, and examined the temporal variability between the first and last time-steps for each observation. Seventeen nights were identified as having unacceptably variable gains, with an average of more than $12^\circ$ of phase change between the first and last time-steps of at least one calibrator, a level at which the imaging quality became very poor. These nights were re-observed in the 2019-A semester. In 2020, the COVID-19 pandemic reduced the observing time available in the 2020-A and B semesters in the extended configuration, so at the time of writing, no further observations to replace any other ionospherically disturbed nights have been possible, although further observations have been proposed for 2022. \Tab~\ref{tab:obs} summarises the observations taken over the period 2018--2020, including those nights that were reobserved.

\section{Continuum pipeline}\label{sec:dr}

The GLEAM-X pipeline is available on GitHub\footnote{\url{https://github.com/tjgalvin/GLEAM-X-pipeline}} in a containerised version that can be run on any platform with Singularity installed \citep{2017PLoSO..1277459K}.

Some common software packages are used throughout the data reduction. Unless otherwise specified:
\begin{itemize}
\item To convert radio interferometric visibilities into images, we use the widefield imager \textsc{WSClean} \citep{2014MNRAS.444..606O} version 2.9, which correctly handles the non-trivial $w$-terms of MWA snapshot images; versions 2 onward include some useful features such as automatically-thresholded \textsc{clean}ing, and multi-scale \textsc{clean} \citep{2017MNRAS.471..301O};
\item the primary beam is as defined by \cite{2017PASA...34...62S}; however, for speed, all primary beams are precalculated and then interpolated as required using code which is available on github\footnote{\url{https://github.com/johnsmorgan/mwa\_pb\_lookup}} and archived on Zenodo \citep{john_morgan_2021_5083990};
\item to mosaic together resulting images, we use the mosaicking software \textsc{swarp} \citep{2002ASPC..281..228B}; to minimise flux density loss from resampling, images are oversampled by a factor of four when being regridded, before being downsampled back to their original resolution;
\item to perform source-finding, we use \textsc{Aegean} v2.2.5\footnote{\url{https://github.com/PaulHancock/Aegean}} \citep{2012MNRAS.422.1812H, 2018PASA...35...11H} and its companion tools such as the Background and Noise Estimator (\textsc{BANE}); this package has been optimised for the wide-field images of the MWA, and includes the ``priorised'' fitting technique, which is necessary to obtain flux density measurements for sources over a wide bandwidth. Fitting errors calculated by \textsc{Aegean} take into account the correlated image noise, and are derived from the fit covariance matrix, which quantifies the quality of fitting; if the fit is poor, and the residuals are large, the fitting errors on position, shape, flux density etc all increase appropriately, so it produces useful error estimates for further use.
\end{itemize}

We now discuss the typical steps undertaken by the pipeline to produce a set of continuum images and catalogues.

\subsection{Calibration}\label{sec:calibration}
Calibration is performed separately on each observation in a direction-independent manner.
The sky model is mainly derived from GLEAM, with additional measurements from the literature for the brighter and more complex sources (e.g. Virgo~A in this release). The sky model is described in a companion paper (Hurley-Walker et al. in prep).
\textsc{MitchCal} \citep{2016MNRAS.458.1057O} is used to generate a calibration solution for each observation, using the full time range of two minutes. These calibration solutions consist of a complex gain for all 4 polarisation products (i.e. a Jones matrix) per tile, per (40-kHz) spectral channel. Since the sky model is limited by the resolution of GLEAM, we exclude baselines longer than the maximum baseline of the 128T configuration, i.e., 2.5\,km ($1667\lambda$ at 200\,MHz); to avoid contamination from diffuse Galactic emission, we also exclude baselines shorter than 112\,m ($75\lambda$ at 200\,MHz).
Calibration solutions are inspected for each night, and tiles or receivers are flagged if they show instrumental issues (e.g., phases appear random with respect to frequency). This typically affects between 1 and 8 of 128~available tiles per night.
We also examine whether the solutions are stable within an observation: rapidly changing gains indicate that ionospheric conditions will dramatically reduce imaging quality (as in \Sect~\ref{sec:obs}). Observations in this category are triaged and do not proceed to imaging (\Sect~\ref{sec:images}).
Similarly, the stability of the gains over the night is inspected; in good conditions, the phases of the solutions only change slowly, on the order of $10^\circ$ on timescales of hours.
If more than $20$\,\% of the solutions for a given observation are flagged, we transfer solutions from a well-calibrated observation at the same frequency that is closest in time.

\subsubsection{Removing contamination from sidelobe sources}

The very brightest radio sources in the sky, the so-called ``A-team'' sources \citep[Table~2 in][]{2017MNRAS.464.1146H}, can cause significant image artefacts if they are just outside the field-of-view or in a sidelobe of the primary beam. Additionally, if they are located inside the field-of-view, the standard deconvolution process (\Section~\ref{sec:imaging}) is not always optimal. To remove these sources from the affected observations, we perform a $(u,v)$ subtraction method. The visibilities are phase-rotated to the location of the source, and a $20'\times20'$ image of the region is formed, using the following \textsc{WSClean} settings:
\begin{itemize}
    \item imaging the XX and YY instrumental polarisation products;
    \item each polarisation product is imaged across 64 480-kHz wide channels that are jointly-\textsc{clean}ed using the \texttt{-join-channels} option, which also produces a 30.72-MHz wide multi-frequency synthesis (MFS) for each polarisation;
    \item a fourth-order polynomial via the \texttt{-fit-spectral-pol} argument to constrain the spectral behaviour of each clean component; %, which was used to captured non-physical spectral behaviour introduced by (potential) chromatic behaviours of the MWA primary beam response
    \item automatic thresholding down to 3$\sigma$, where $\sigma$ is measured as the root mean square (RMS) of the residual XX and YY MFS images at the end of each major cycle;
    \item a major clean cycle gain of 0.85, i.e. removes 85\% of the flux density of the clean components at the end of each major cycle;
    \item ``Briggs'' \citep{Briggs1995} robust parameter of $-1$;
    \item 10 or fewer major clean cycles, in which the images are inverse Fourier transformed back to visibilities, which are subtracted from the data;
    \item Up to $10^5$ minor cleaning cycles, where the subtraction takes place in the image plane.
\end{itemize}

During this process the ``MODEL'' column of the measurement set\footnote{Measurement sets are a common form of data structure for radio interferometric data. They typically consist of three columns: the ``DATA'', usually produced directly by the instrument; ``CALIBRATED\_DATA'' to which some calibration has been applied; and ``MODEL'', the model sky sampled by the instrument, which is typically produced either during calibration or imaging.}  is updated with the source components, and after it has completed, is subtracted from the calibrated visibilities. The observation is then phase-rotated back to its original location. In this way, the chromatic effect of the primary beam sidelobe is taken into account when removing the source, without distorting the overall gains of the observation.\footnote{This differs from the peeling approach used by \cite{2017MNRAS.464.1146H} to process the GLEAM data, which was found to require too much manual intervention for a survey of the size of GLEAM-X.}

\subsubsection{Polarisation calibration}

We also introduce two extra steps in the calibration stage to make the measurement sets ready for polarisation analysis. One is the parallactic angle correction within the primary beam model \citep{2017AJ....154...54H}, transforming the data from the observed frame (linear feeds on the ground) to an astronomical reference frame according to the IAU standard (polarisation angle measured from North through East). This step is necessary for linear polarisation analysis when observations cover a large range of hour angles.
To facilitate later polarisation analysis, we set the cross-terms of the calibration Jones matrices to zero, as well as dividing the Jones matrices for all tiles through by a phasor representing the phases of a reference antenna, which is used for all survey processing. At the same time, we add an X-Y phase determined from observations of a strong polarised source with a known polarisation angle \citep{2017PASA...34...40L}. 
The X-Y phase correction reduces the leakage between linear and circular polarisation, making circularly polarised data available.  A detailed description of polarisation calibration, imaging, and a first data release will be given in a separate publication describing the POlarised GLEAM-X survey (POGS-X; Zhang et al. in prep).

\subsection{Imaging}\label{sec:imaging}

At this stage, the processing diverges depending on whether there is significant Galactic emission. For this paper, we focus on producing catalogues and images which best explore the extragalactic sky (i.e. without attempting to reconstruct such diffuse emission).

\begin{figure}[t]
    \centering
    \includegraphics[width=1\linewidth]{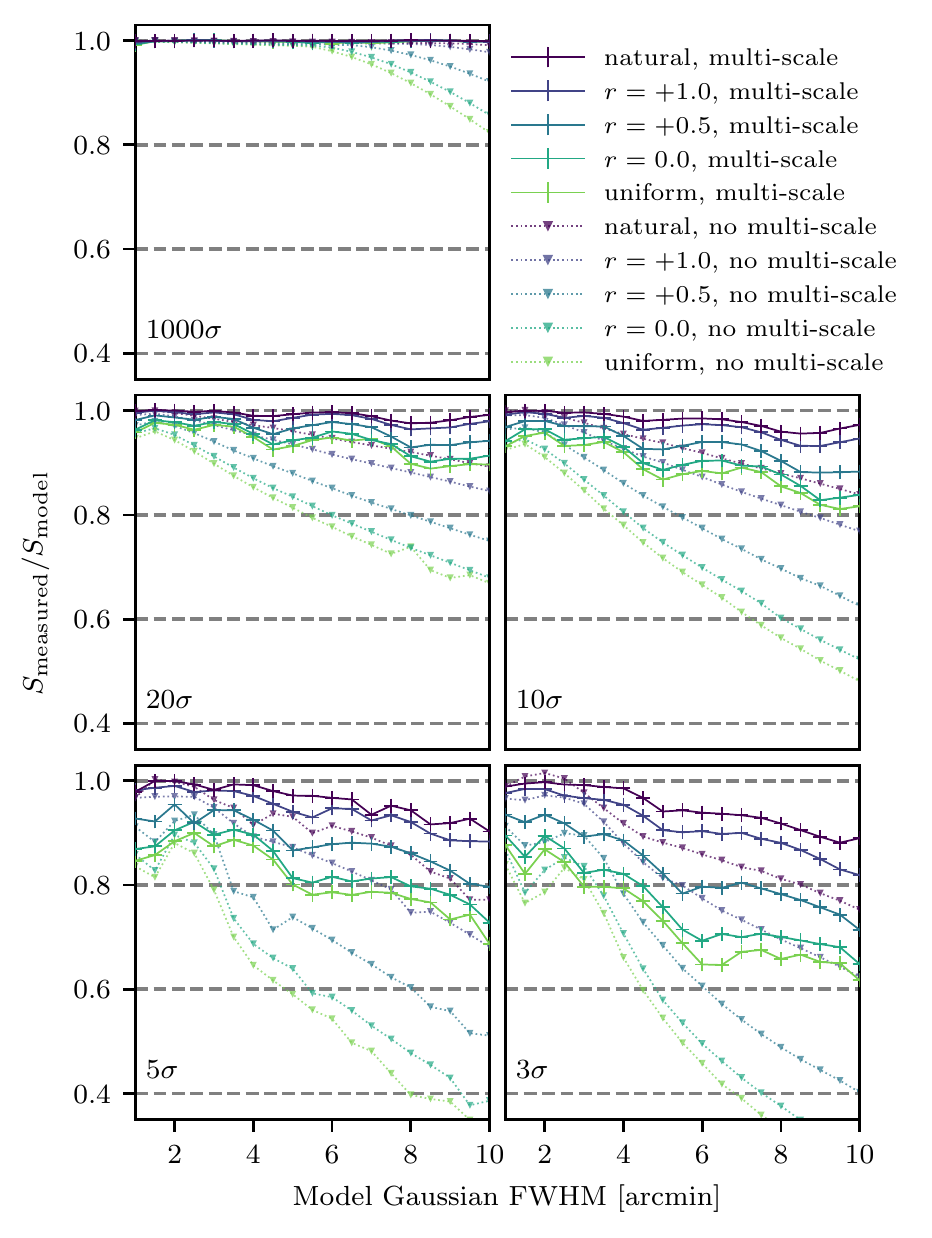}
    \caption{\label{fig:fluxrecovery} Flux density recovery fraction for model 3, 5, 10, 20, and $1000\sigma$ 2-d Gaussian sources with varying FWHM for a range of image weightings. The FWHM range from 60~arcsec up to 600~arcsec in 30~arcsec intervals. The PSF major axis FWHM varies from 80~arcsec (for uniform weighting) to 115~arcsec (for natural weighting). Note the step-pattern visible in the multi-scale cases likely arises due to choice of scales during multi-scale CLEAN.}
\end{figure}

While the original GLEAM survey used an image weighting with a ``Briggs'' robust parameter $-1$, such a  weighting is not suitable for the MWA Phase~\textsc{II} extended configuration, as the latter has fewer short baselines, reducing the surface brightness sensitivity. For GLEAM-X, a weighting closer to natural is generally preferred to maximise sensitivity \citep[see][for a demonstration of the surface brightness sensitivity of MWA Phase~\textsc{II} in comparison to other instruments]{2020PASA...37...32H}.

To determine an appropriate weighting for extended MWA Phase~\textsc{II} imaging, taking into account both angular resolution and surface brightness sensitivity, we trial a range of image weightings, including ``Briggs'' weighting with robust parameters $0.0$, $+0.5$, and $+1.0$, as well as uniform and natural weightings. We simulate simple 2-dimensional Gaussian sources with varying full-width at half-maximum (FWHM) in individual template 154-MHz 2-min snapshots after subtracting astronomical sources and noise. Two runs of normal snapshot imaging are performed for each Gaussian source---one with multi-scale CLEAN enabled and the other without. The flux density of the resultant Gaussian sources was then measured using the source-finding software \textsc{aegean} to model the Gaussian component. For the purpose of simulating and measuring the model sources at 3, 5, 10, 20, and 1000$\sigma$, and an RMS noise level $\sigma$ is estimated from real template images for the given image weightings.

Fig.~\ref{fig:fluxrecovery} shows the various image weightings for the imaging with/without multi-scale CLEAN with the \textsc{aegean} flux density measurements of the sources. A significant increase in the recovered flux density during multi-scale CLEAN motivates its use. The `best' case for flux density recovery is a natural weighting with multi-scale CLEAN, however with natural weighting the improvement in angular resolution compared to GLEAM is only a factor of $\sim 1.5$ and the point source sensitivity is not maximised. To balance an increase in resolution while retaining overall sensitivity, a ``Briggs'' robust parameter of $+0.5$ is chosen for the full survey. We note that the fraction of flux density loss decreases with increasing source brightness. For instance, comparing the top and leftmost two panels of Fig.~\ref{fig:fluxrecovery}, 90\,\% of the flux density is recovered for a $10'$-FWHM 20-$\sigma$ source, whereas all of the flux density would be recovered for a 1000-$\sigma$ source of the same size. 

While these simulations provide an estimate of the flux density recovery for extended Gaussian sources in snapshot observations, the results shown in Fig.~\ref{fig:fluxrecovery} should not be used to directly correct flux density measurements made in the final mosaics.  

\textsc{WSClean} is used to generate images with the following settings:
\begin{itemize}
\item A SIN projection centred on the minimum-$w$ pointing, i.e. hour angle~$=0$, Dec~$-26.7^\circ$
\item four 7.68-MHz channels jointly-cleaned using the \texttt{-join-channels} option, which also produces a 30.72-MHz MFS image;
\item include and apply the MWA primary beam \citep{2017PASA...34...62S} during cleaning, to produce a Stokes~I image;
\item automatic thresholding down to $3\sigma$, where $\sigma$ is the RMS of the residual MFS image at the end of each major cycle;
\item automatic \textsc{clean}ing down to 1$\sigma$ within pixels identified as containing flux density in previous cycles (``masked'' \textsc{clean}ing);
\item a major cycle gain of 0.85, i.e. 85\% of the flux density of the clean components are subtracted in each major cycle;
\item five or fewer major cycles, in order to prevent the occasional failure to converge during cleaning between 3 and 4$\sigma$;
\item $10^6$ minor cycles, a limit which is never reached;
\item \texttt{multiscale} \textsc{Clean}, with the default deconvolution scale settings, and a multiscale-gain parameter of 0.15;
\item $8000\times8000$ pixel images, which encompasses the field-of-view down to 10\% of the primary beam;
\item ``robust'' weighting of 0.5 (see above); 
\item a frequency-dependent pixel scale such that each image always has 3.5--5 pixels per FWHM of the restoring beam;
\item a restoring beam of a 2-D Gaussian fit to the central part of the dirty beam, which is similar in shape (within 10\,\%) for each frequency band of the entire survey, but varies in size depending on the frequency of the observation.
\end{itemize}

\begin{figure*}
    \centering
    \includegraphics[width=\textwidth]{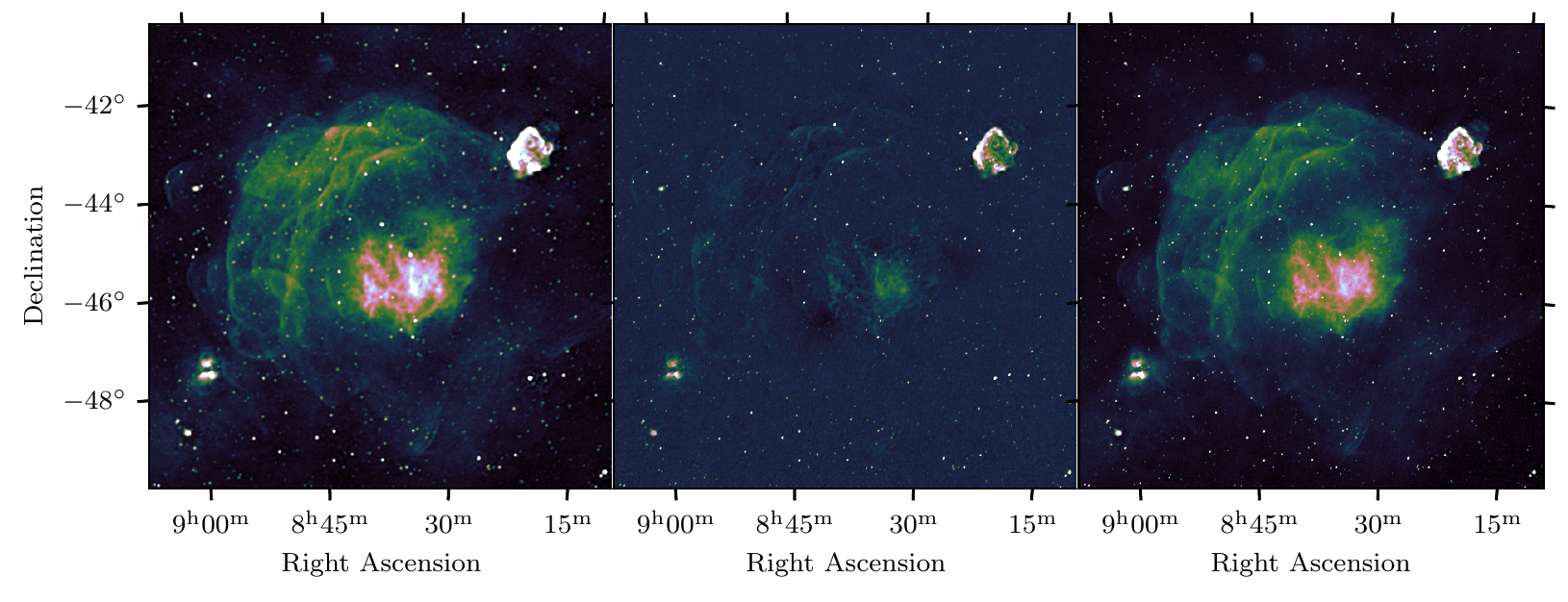}
    \caption{90\,sq.\,deg. around the Vela supernova remnant at 139--170\,MHz. The left panel shows a GLEAM mosaic at 2\farcm6 resolution; the middle panel shows a GLEAM-X mosaic at 1\farcm3 resolution; the right panel shows a joint deconvolution of the two datasets yielding the same high resolution, and also the sensitivity to structures on $10'$--$5^\circ$ scales.}
    \label{fig:Vela}
\end{figure*}

The extended configuration of the Phase~\textsc{II} MWA has low sensitivity to sources with extents $>10'$, and thus is not optimal for recovering the complex emission present in the Galactic Plane. However, the original GLEAM survey was recorded in an identical set of drift scan pointings to GLEAM-X, and at that time the array configuration provided many baselines with sensitivity to these larger angular scales. Thus, for the Galactic plane, we will jointly deconvolve the short baselines of GLEAM with the full GLEAM-X measurement sets, a process enabled by the fast GPU-based image-domain gridding extension to \textsc{WSClean} \citep{2018A&A...616A..27V}. This method has been used to great effect to image Fornax~A \citep{2020PASA...37...27L} and Centaurus~A \citep{2022NatAs...6..109M}, and can also be used for other extended sources such as the Magellanic Clouds. An example of these results is shown in \Fig~\ref{fig:Vela} and the full description of the process in the context of the Galactic Plane will be demonstrated in a further paper (Hurley-Walker et al. in prep).

\subsection{Astrometric calibration}\label{sec:astrometry}

The ionosphere introduces a $\lambda^2$-dependent position shift to the observed radio sources, which varies with position on the sky. Following the method of \cite{2017MNRAS.464.1146H} and \cite{2019PASA...36...47H}, we use \textsc{fits\_warp} \citep{2018A&C....25...94H} to calculate a model of position shifts based on the difference in positions between the sources in the snapshot and those in a reference catalogue, and then use this model to de-distort the images. 

For the reference catalogue, we benefit from using catalogues with similar resolution ($\sim$1$'$) covering wide areas. For declinations north of $-30^\circ$, we use NVSS at 1.4\,GHz, and for the southern sky SUMSS at 843\,MHz. From this combined catalogue we select a subset which is sparse (no internal matches within $3'$) and unresolved (integrated to peak flux density ratio of $<1.2$).

For each of the 7.68-MHz sub-bands and the wideband 30.72-MHz images formed from each observation, we estimate the background and RMS noise $\sigma$ using \textsc{BANE}, and perform source-finding using \textsc{Aegean}, with a minimum ``seed'' threshold of $5\sigma$. Using the iterative catalogue cross-matching functionality of \textsc{fits\_warp}, we cross-match the measured sources to the reference catalogue, typically finding 1,000--3,000 cross-matches, from which we retain the 750~brightest sources. A greater number of sources does not improve the accuracy of the warping for typical ionospheric conditions, but does add computational load, so this value was chosen as a point of diminishing returns. These sources typically have flux densities $>1$\,Jy in the NVSS and SUMSS surveys so have adequate astrometry to form the baselines for our corrections.

Snapshot images with fewer than 100~successful cross-matches are discarded (typically $<1$\,\% of images). The position shifts in the remaining images are typically of order $25''$--$5''$ over 72--231\,MHz, and are coherent on scales of 1--20$^\circ$, similar to previous studies with the MWA \citep[e.g.][]{2018A&C....25...94H,2020RaSc...5507106H}. \textsc{fits\_warp}  uses these position shifts to create a warp model, apply it to all pixels, and interpolate the results back on to the original pixel grid. This technique yields residual astrometric offsets (with no obvious preferred direction or structure) of order $6''$ at the lowest frequencies, and $2''$ at the highest frequencies.

\subsection{Primary beam correction}\label{sec:pbcorr}

While the primary beam model developed by \cite{2017PASA...34...62S} is significantly more accurate than previous models of the MWA primary beam, there remain some discrepancies between our measured source flux densities and those predicted from existing work. In part, this is due to the flagging of individual dipoles in different tiles across the array, which gives these tiles a different and unmodelled primary beam response. For the observations processed in this work, 72 tiles were fully functional, 39~tiles contained one dead dipole, 14 contained two dead dipoles, and three tiles were flagged for having three or more non-functional dipoles. Including the effect of the flagging by computing and using multiple primary beams at the calibration and imaging stages is computationally expensive, so instead a correction is made after the images are formed.

 We cross-match each snapshot with a sparse (no internal matches within $5'$), unresolved (major axis $a\times$ minor axis $b<2'\times2'$) version of the GLEAM-derived catalogue used for calibration (\Sect~\ref{sec:calibration}) and make a global mean flux density scale correction using the \textsc{flux\_warp}\footnote{\url{https://gitlab.com/Sunmish/flux_warp}} package \citep{2020PASA...37...37D}, typically of order 5--15\,\%.
 After this global shift, we accumulate the cross-matched tables. Since the discrepancy is consistent in Hour Angle and Dec between snapshots, we can combine the information in this frame of reference.

\begin{figure}
    \centering
    \includegraphics[width=1\linewidth]{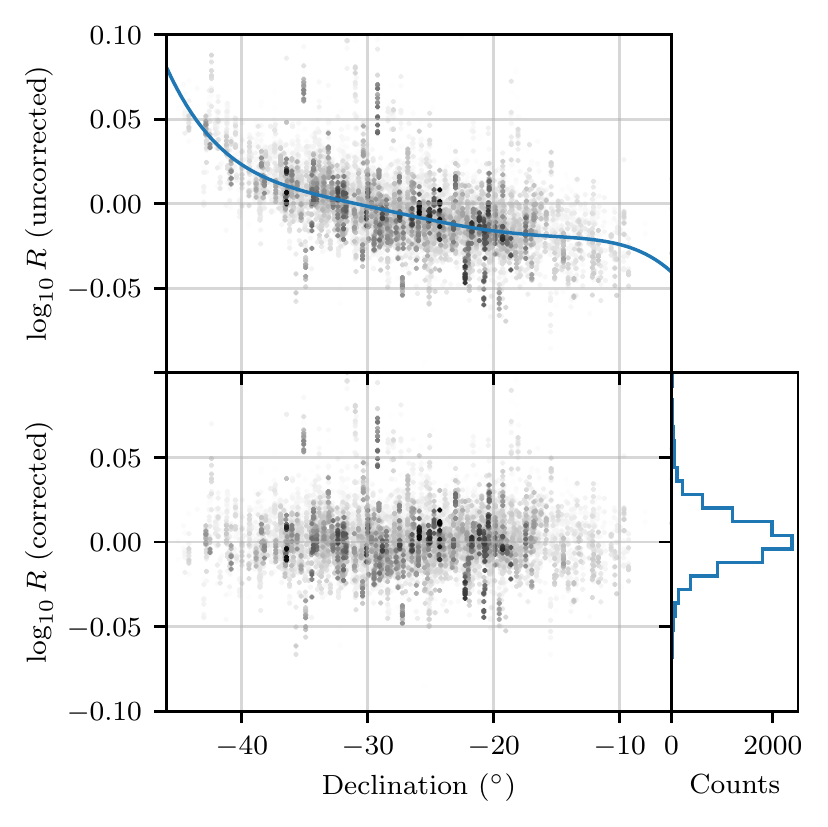}
    \caption{The accumulated flux density scale distribution across the snapshot images at 147--155\,MHz from observations performed on 2018-02-20. The upper panel shows the change in $\log_{10}R$ as a function of Dec, where $R$ is the ratio of measured integrated flux density to model integrated flux density. The lower panel shows the same after the polynomial correction function (blue line) has been fit and applied. The adjacent histogram shows the resulting distribution of $\log_{10}R$ over the drift scan. The full-width-at-half-maximum of the resulting histogram is $\sim2.5$\,\%. Similar results are obtained for other frequency bands.
    \label{fig:drift_poly}}
\end{figure}

For each frequency, as a function of HA and Dec, we compare the $\log_{10}$ of the ratio $R$ of the integrated flux densities of the measured source values and reference catalogue.
Similarly to GLEAM ExGal, we find no trends with HA, and up to $\pm10$\,\% trends in Dec. \Fig~\ref{fig:drift_poly} shows this effect for a typical frequency band.
A fifth-order polynomial model is fitted as a function of Dec using a weighted least-squares fit, where the weights are the signal-to-noise of the sources as measured in each snapshot. The standard deviation of the data from the model
($\sigma_\mathrm{poly}$) is measured,
and sources with $|R|>3\times\sigma_\mathrm{poly}$ are removed from the data. A final model of the same form is fitted to the remaining data, forming a correction function which is then applied to every individual snapshot.

After correction, the primary-beam-corrected 30-MHz MFS images have snapshot RMS values of 35--4\,mJy\,beam$^{-1}$ over 72--231\,MHz at their centres, where the primary beam sensitivity is highest.

\subsection{Mosaicking}\label{sec:mosaic}

The goal of continuum mosaicking is to combine the astrometrically- and primary-beam-corrected snapshot images into deeper images with reduced noise, revealing fainter sources and diffuse structures invisible in the individual snapshots.
For optimal signal-to-noise when mosaicking the night-long scans together, we use inverse-variance weighting. The weight maps are derived from the square of primary beam model, scaled by the inverse of the square of the RMS of the center of the image, as calculated by BANE.

As discussed in \Sect~\ref{sec:obs}, GLEAM-X was observed at three different hour angles. This gives each drift scan slightly different $(u,v)$-coverage, which results in a slightly different restoring beam and thus point spread function (PSF). While each individual drift scan would have a unique and very nearly Gaussian PSF, it could be expected that a stacking of different unique Gaussians with different position angles would result in a non-Gaussian shape. Since most source-finders expect sources to be well-approximated by Gaussians, we tested this effect in our mosaicking procedure. We selected the scans with the most dissimilar $(u,v)$-coverage where there would be significant overlap in sources, those at HAs $-1$ and $+1$ from the Dec$+2$ scans, i.e. where the sky is rotating most quickly and projection effects are most important. We simulated a grid of 1~Jy point sources at common RA and Decs for seven observations from each of these scans, and ran them through our imaging and mosaicking stages, using unity image weighting and neglecting unnecessary astrometric and primary beam corrections. We used \textsc{Aegean} to source-find on the resulting mosaic, making corrections as necessary for the projection (\Sect~\ref{sec:catalogue}). We recovered the sources at integrated flux densities of 0.995--0.999\,Jy and peak flux densities of 0.96--0.97\,Jy\,beam$^{-1}$. Subtracting these Gaussian fits from the image plane data, we found residuals at the $<4.5$\,\% level, indicating that level of deviation away from Gaussianity. Since the integrated flux densities were recovered well, and the non-Gaussianity is fairly small, even for this worst-case scenario, we adopt this mosaicking method going forward.

For each 7.68-MHz frequency channel, we form a night-long drift scan, and examine it to check for any remaining data quality issues. We also form five~30.72-MHz bandwidth mosaics from the multi-frequency synthesis images generated during cleaning (\Sect~\ref{sec:imaging}). After quality checking, for each frequency, data from all four nights that cover the same RA range are combined together to make a single deep mosaic.
At this stage, we also form a 60-MHz bandwidth ``wideband'' image over 170--231\,MHz, as this gives a good compromise between sensitivity and resolution, and will be used for source-finding (\Sect~\ref{sec:catalogue}).

\subsection{Calculation of the PSF}\label{sec:psf}

As described in Appendix~A of \citet{2020PASA...37...37D}, imaging away from the phase centre incurs a significant phase rotation during re-gridding as part of the mosaicking process. This re-projection results in a point-spread function that is not defined at the image reference coordinates.  This is corrected partially by introduction of a projected regrid factor, $f$, that is applied to the PSF major axis to form an `effective' PSF major axis. For a resultant ZEA projection this is simply related to the change in solid angle over the original SIN-projected image with \citep[e.g.][]{TMS} \begin{equation}
    f = \sqrt{1 - l^2 -m^2} \, ,
\end{equation}
where $l$ and $m$ are the direction cosines defined with reference to the \textit{original}, SIN-projected image direction. The ZEA projection itself reduces additional area-related projection effects due to its equal area nature. This is used in initial source-finding on the mosaics as the integrated flux density is correct and the product of the major and minor PSF axes is also correct for the new projection.

Residual uncorrected ionospheric distortions can cause slight blurring of the final mosaicked PSF. This can be characterised by examining sources which are known to be unresolved, which is best determined by using a higher-resolution catalogue than our calibration sky model; we thus use the NVSS and SUMSS combined catalogue described in \Section~\ref{sec:astrometry}.
Following \cite{2017MNRAS.464.1146H,2019PASA...36...47H}, we cross-match this catalogue with the sources detected in our mosaics at signal-to-noise$>10$, and then measure the size and shape of these sources in the GLEAM-X mosaics. We create a PSF map by averaging and interpolating over these sources, using Healpix (order$=4$, i.e. pixels $\sim$3$^\circ$ on each side) as a natural frame in which to accumulate and average source measurements.

After the PSF map has been measured, its antecedent mosaic is multiplied by a (position-dependent) ``blur'' factor of
\begin{equation}
\centering
B = \frac{a_{\mathrm{PSF}} b_{\mathrm{PSF}}}{a_{\mathrm{rst}} b_{\mathrm{rst}}}
\label{eq:ionoblur}
\end{equation}

\noindent where $a_\mathrm{rst}$ and $b_\mathrm{rst}$ are the FWHM of the major and minor axes of the restoring beam, and $a_\mathrm{PSF}$ and $b_\mathrm{PSF}$ are the FWHM of the major and minor axes of the PSF. This has the effect of normalising the flux density scale such that both peak and integrated flux densities agree, as long as the correct, position-dependent PSF is used \citep{2018PASA...35...11H}. Values of $B$ are typically 1.0--1.2.

\subsection{Final images}\label{sec:images}

\begin{sidewaysfigure*}
    \centering
    \includegraphics[width=\textwidth]{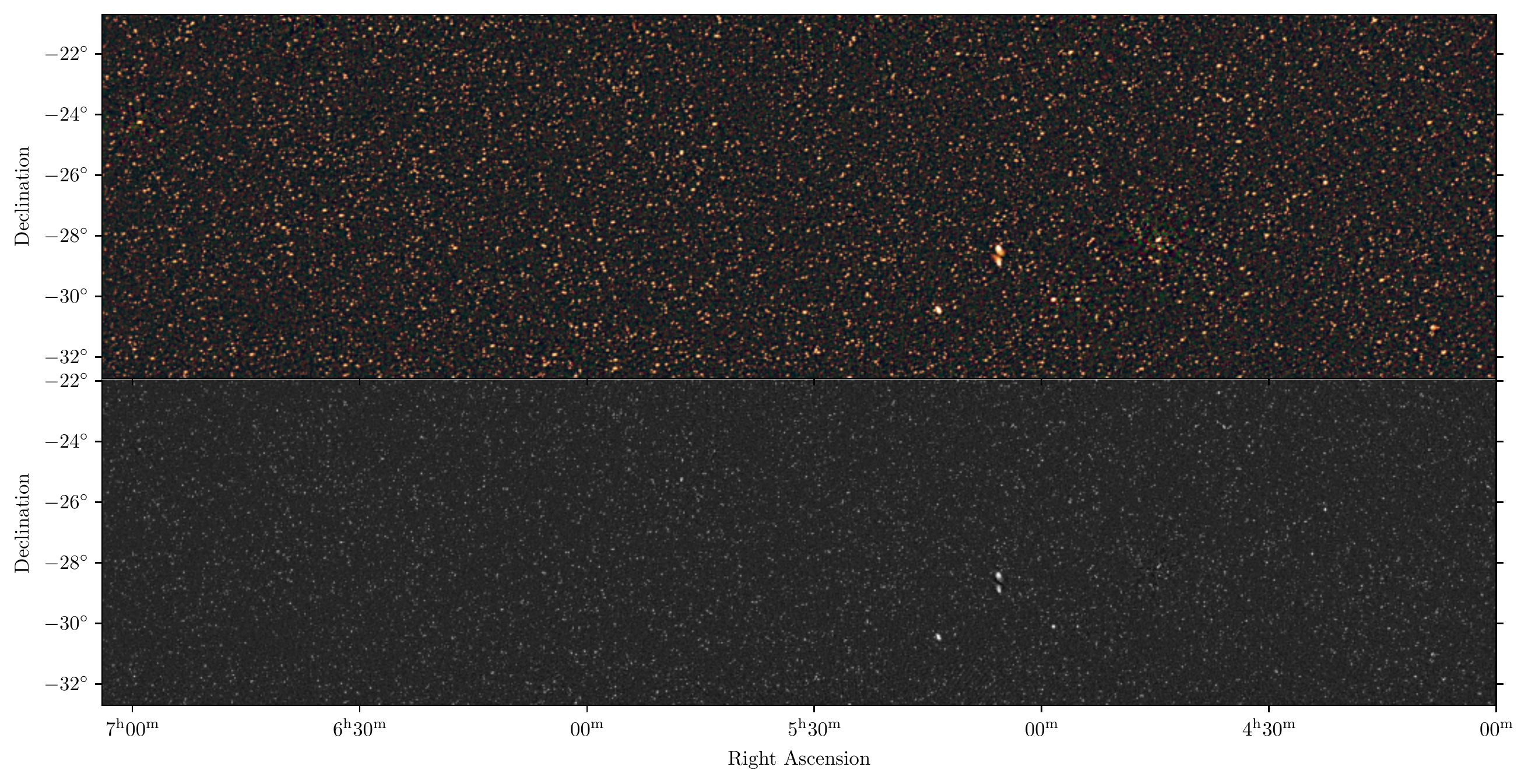}
    \caption{Continuum mosaics from this data release, for RA 4--7\,h. In each sub-panel, the top image shows the 72--103\,MHz (R), 103--134\,MHz (G), and 139--170\,MHz (B) data as an RGB cube, with an $\arcsinh$ stretch spanning $-9$--200\,mJy\,beam$^{-1}$; the lower image shows the 170--231\,MHz source-finding image, with an $\arcsinh$ stretch spanning $-3$--200\,mJy\,beam$^{-1}$.}
    \label{fig:headline1}
\end{sidewaysfigure*}

\begin{sidewaysfigure*}
    \centering
    \includegraphics[width=\textwidth]{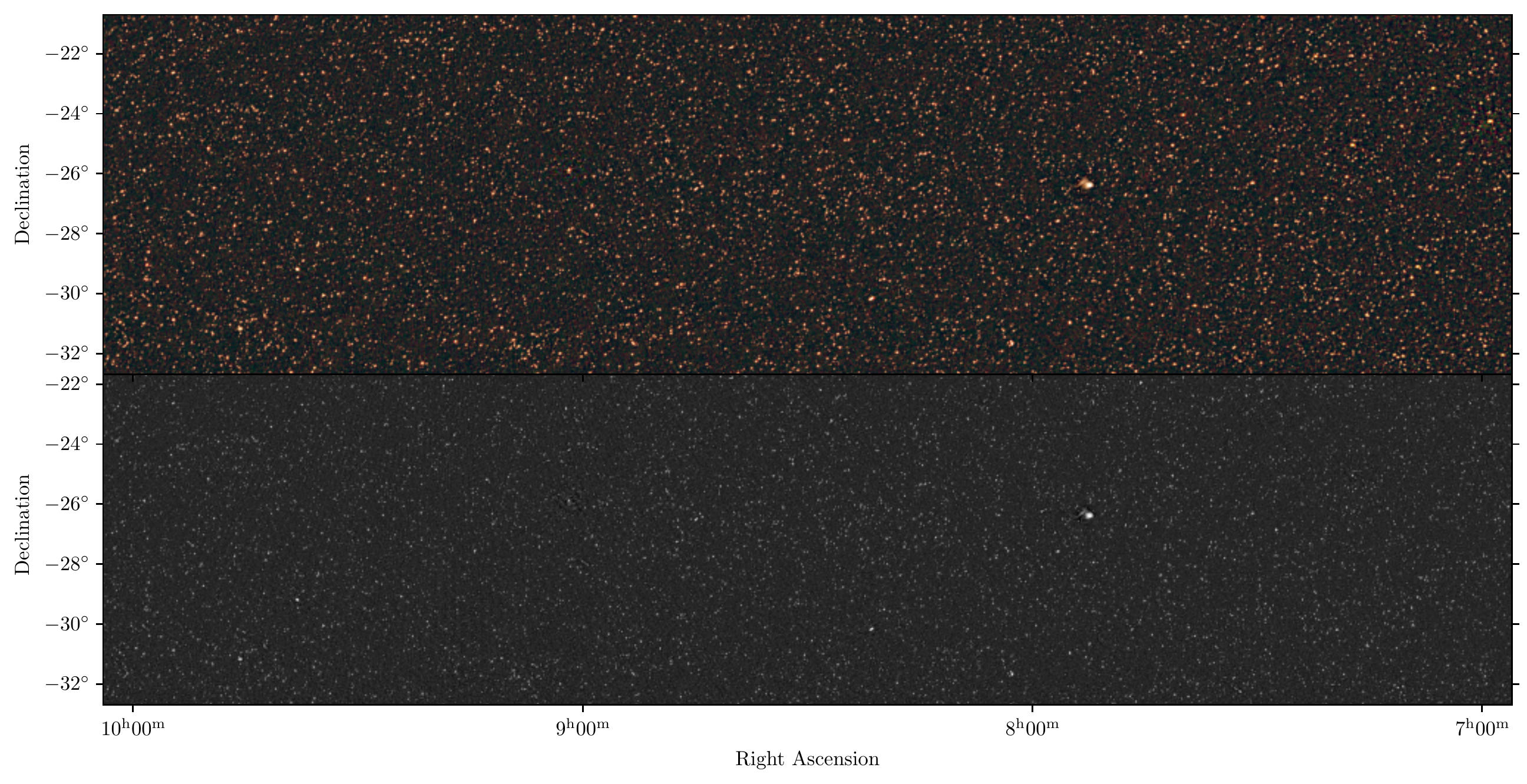}
    \caption{Continuum mosaics from this data release, for RA 7--10\,h. Figure formatting is identical to \Fig~\ref{fig:headline1}.}
    \label{fig:headline2}
\end{sidewaysfigure*}

\begin{sidewaysfigure*}
    \centering
    \includegraphics[width=\textwidth]{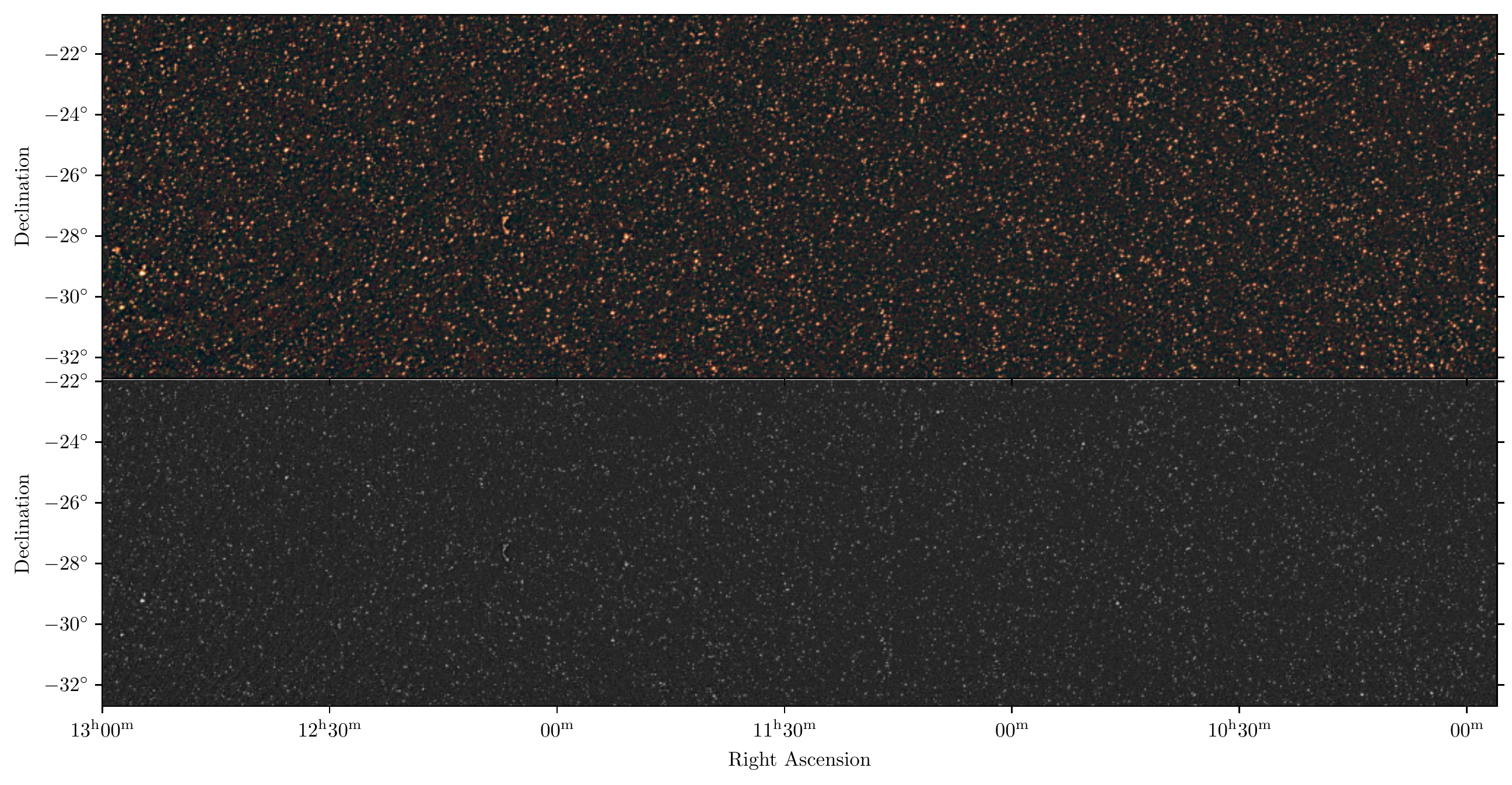}
    \caption{Continuum mosaics from this data release, for RA 10--13\,h. Figure formatting is identical to \Fig~\ref{fig:headline1}.}
    \label{fig:headline3}
\end{sidewaysfigure*}

The mosaicking stage of \sect~\ref{sec:mosaic} results in 26~mosaics: one with 60-MHz bandwidth, five with 30-MHz bandwidth and the other~20 covering 72--231MHz in 7.68-MHz narrow bands.
In this work, we run the pipeline on four nights of observing indicated in \Tab~\ref{tab:obs}, producing a large set of mosaics with decreasing sensitivity toward the edges.
Here we downselect to a region which is representative of the survey's eventual sensitivity, covering 4\,h$\leq$ RA$\leq 13$\,h, $-32.7^\circ \leq \mathrm{Dec} \leq -20.7^\circ$, for further analysis. \Figs~\ref{fig:headline1}--\ref{fig:headline3} show this area for four of the deeper mosaics. Postage stamps of these images are available on both SkyView and the GLEAM-X website\footnote{\url{https://www.mwatelescope.org/gleam-x}}. The header of every postage stamp contains the PSF information calculated in \Sect~\ref{sec:psf}, and the completeness information calculated in \Sect~\ref{sec:reliability}. 

%\begin{figure*}
%    \centering
%    \includegraphics[width=0.8\textwidth]{figs/headline_gleamx.pdf}
%    \caption{Continuum mosaics from this data release. In each sub-panel, the top image shows the 72--103\,MHz (R), 103--134\,MHz (G), and 139--170\,MHz (B) data as an RGB cube, with an $\arcsinh$ stretch between $-9$--200\,mJy\,beam$^{-1}$; the lower image shows the 170--231\,MHz source-finding image, with an $\arcsinh$ stretch between $-3$--200\,mJy\,beam$^{-1}$.}
%    \label{fig:headline}
%\end{figure*}

% TopCat for getting 548 sources for figure below:
% ra<159.25&&ra>155.75&&dec<-25.75&&dec>-29.25&&(int_flux/local_rms>=5)

\begin{figure*}
    \centering
    \includegraphics[width=\textwidth]{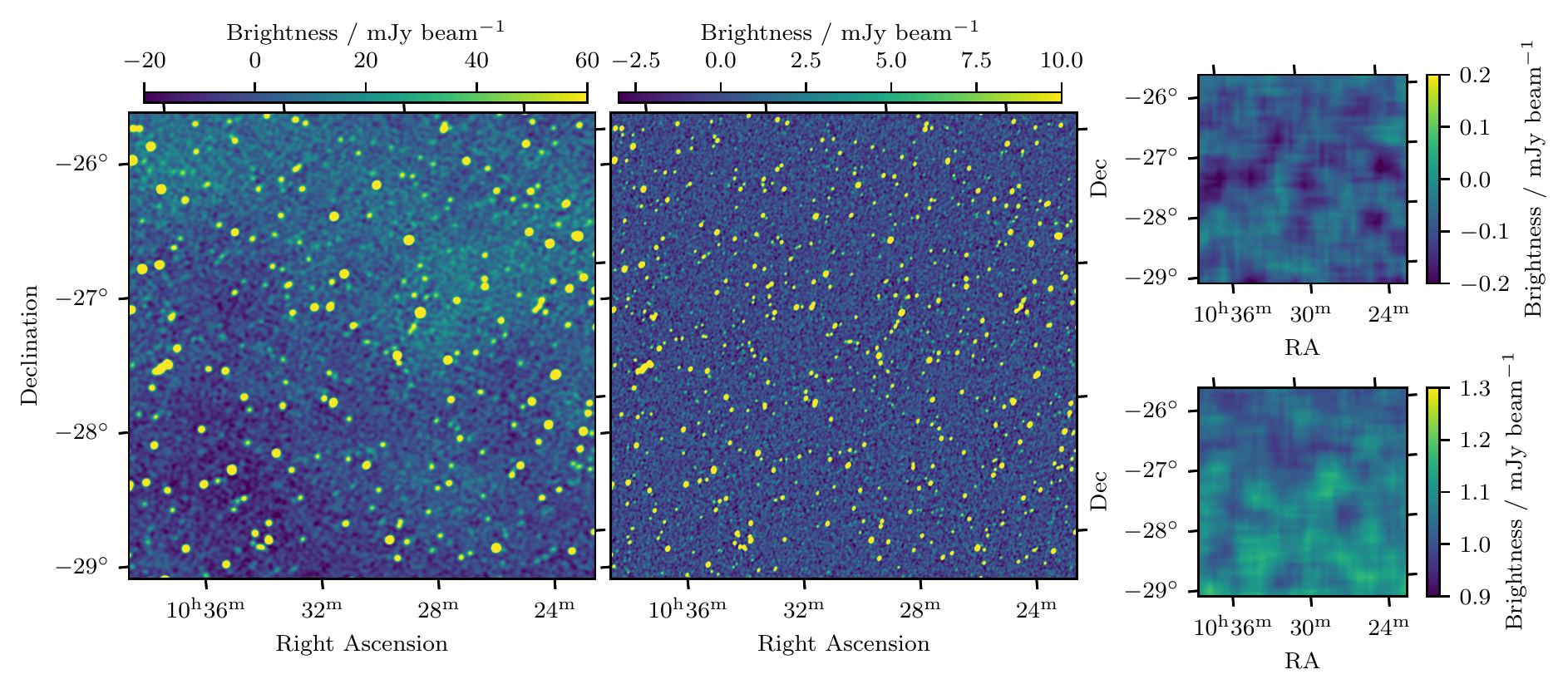}
    \caption{Ten~sq.~deg. of the wideband source-finding mosaic centered on RA~$10^\mathrm{h}30^\mathrm{m}$ Dec$-27^\circ30'$; the left panel shows the image from GLEAM ExGal; the middle panel shows the image from this work; the top and bottom right panels show, respectively, the background and RMS noise of the GLEAM-X data as measured by \textsc{BANE}. GLEAM ExGal contains 212~components in this region, and the average RMS noise is 6\,mJy\,beam$^{-1}$; GLEAM-X contains 548~components and the average RMS noise is 1.1\,mJy\,beam$^{-1}$.}
    \label{fig:example_mosaic}
\end{figure*}

We use \textsc{BANE} to determine the background and RMS noise of each mosaic. During development of this survey, we noticed that \textsc{BANE}'s default of three loops of 3-sigma-clipping is insufficient to exclude source-filled pixels to accurately determine the background and RMS noise. The issue may not have been noticed in previous works due to the relatively higher sensitivity and resulting source density of GLEAM-X (although \cite{2017MNRAS.464.1146H} noted a similar effect from the high sidelobe confusion levels of GLEAM). We modified \textsc{BANE} to use 10~loops and found that it produced more accurate noise and background estimates (see \Sect~\ref{sec:noise_properties} for further analysis). \Fig~\ref{fig:example_mosaic} shows an example of 10~sq.~deg of the 170--231\,MHz wideband mosaic and associated background and RMS noise, as well as the same region as seen by GLEAM ExGal, in which the resolution is lower, the noise is higher, and the diffuse Galactic synchrotron on scales of $>1^\circ$ is visible.

Combining data in the image plane may lead to the recovery of faint sources that were not cleaned during imaging. The RMS noise levels in the wide-band (30-MHz) mosaics range from 5--1.3\,mJy\,beam$^{-1}$ over 72--231\,MHz. This compares to typical snapshot RMS values of 35--4\,mJy\,beam$^{-1}$ over the same frequency range (\Sect~\ref{sec:pbcorr}). Cleaning is performed down to 1-$\sigma$ for components detected at 3-$\sigma$ in a snapshot (\Sect~\ref{sec:imaging}). The centres of each image form the greatest contribution to each mosaic due to weighting by the square of the primary beam (\Sect~\ref{sec:mosaic}). We can therefore estimate at what signal-to-noise threshold uncleaned sources will typically appear: $\frac{3\times35}{5}=21$ -- $\frac{3\times4}{1.3}=9$ from 72--231\,MHz, and at $\sim\frac{3\times4}{1}=12$ in the wideband (60-MHz) source-finding image (\Sect~\ref{sec:catalogue}). 

Modelling this effect, especially in conjunction with Eddington bias (see e.g. \Sect~\ref{sec:gleam}), which is also significant at these faint flux densities, lies beyond the scope of this paper. It would involve significant work and is mainly of interest for performing careful measurement of low-frequency source counts \citep[see ][for an equivalent analysis for GLEAM]{2019PASA...36....4F}. At this stage we merely suggest additional caution when using flux densities for sources at low ($<12$) signal-to-noise.

The mosaics at this stage are only a subset of the GLEAM-X sky. The RMS increases toward the edges due to the drop in primary beam sensitivity and selected RA range of these observations. Future mosaics comprised of further nights of observing will be combined to produce near-uniform sensitivity across the sky.

\section{Compact Source Catalogue}\label{sec:catalogue}

A source catalogue derived from the images is a useful data product that enables straightforward cross-matching, spectral fitting, and population studies. We aim here to accurately capture components of sizes $<10'$ across all frequency bands, fitting elliptical two-dimensional Gaussians with \textsc{Aegean}. We carry out this process on the \survarea{}-deg$^2$ region selected in \sect~\ref{sec:mosaic}, and the steps are generally applicable to future mosaics produced from the survey.

\subsection{Source detection}\label{sec:detection}

We follow the same strategy as \cite{2017MNRAS.464.1146H}: using the 170--231\,MHz image, a deep wideband catalogue centred at 200\,MHz is formed. We set the ``seed'' clip to four, i.e. pixels with flux density $>4\sigma$ are used as initial positions on which to fit components, where $\sigma$ denotes the local RMS noise. After the sources are detected, we filter to retain only sources with integrated flux densities $\geq5\sigma$.
We then use the ``priorised'' fitting technique to measure the flux densities of each source in the narrow-band images: the positions are fixed to those of the wide-band source-finding image, the shapes are predicted by convolving the shape in the source-finding image with the local PSF, and the flux density is allowed to vary. Where the sources are too faint to be fit, a forced measurement is carried out. 
We perform several checks on the quality of the catalogue, detailed below.

\subsection{Error derivation}

In this \sect{} we examine the errors reported in the catalogue. First, we examine the systematic flux density errors; then, we examine the noise properties of the wide-band source-finding image, as this must be close to Gaussian in order for sources to be accurately characterised, and for estimates of the reliability to be made, which we do in \Sect~\ref{sec:reliability}. Finally, we make an assessment of the catalogue's astrometric accuracy. These statistics are given in \Tab~\ref{tab:survey_stats}.

\subsubsection{Comparison with GLEAM}\label{sec:gleam}

GLEAM forms the basis of the flux density calibration in this work, and in this \sect{} we examine any differences between the flux densities measured here compared to those measured by GLEAM ExGal. We select compact sources from both catalogues (integrated / peak flux density $<2$) that cross-match within a $15''$ radius, and have a good power-law spectral index fit (reduced $\chi^2\leq1.93$; see \Sect~\ref{sec:alpha}). Curved- and peaked-spectrum sources comprise only a small proportion of the catalogue and are more likely to be variable \citep{2021MNRAS.501.6139R}, so are not included in this check. We excluded all sources in GLEAM-X data which have a cross-match within $2'$ in order to avoid selecting sources which are unresolved in GLEAM and resolve into multiple components in GLEAM-X.

As surveys approach their detection limit, measured source flux densities are increasingly likely to be biased high due to noise; there are a larger number of faint sources available to be biased brighter by noise than there are bright sources available to be biased dimmer. \cite{1913MNRAS..73..359E} describes corrections that can be made to an ensemble of measurements to remove this bias. For the purpose of this section, we wish to correct the individual GLEAM flux density measurements in order to check the GLEAM-X flux density scale. We use Eq.~4 of \cite{1998PASP..110..727H} to predict the maximum likelihood true flux density of each of the GLEAM 200-MHz measurements:
\begin{equation}
    S_\mathrm{corrected} = \frac{S}{2} \left( 1 + \sqrt{\frac{4q+4}{\frac{S}{\sigma}^2}}\right)
\end{equation}
where $\sigma$ is the local RMS noise, and $q$ is the logarithmic source count slope (i.e. the index in $\frac{dN}{dS}\propto S^q$); at these flux density levels $q=1.54$ \citep{2016MNRAS.459.3314F}.

\begin{figure}
    \centering
    \includegraphics[width=1\linewidth]{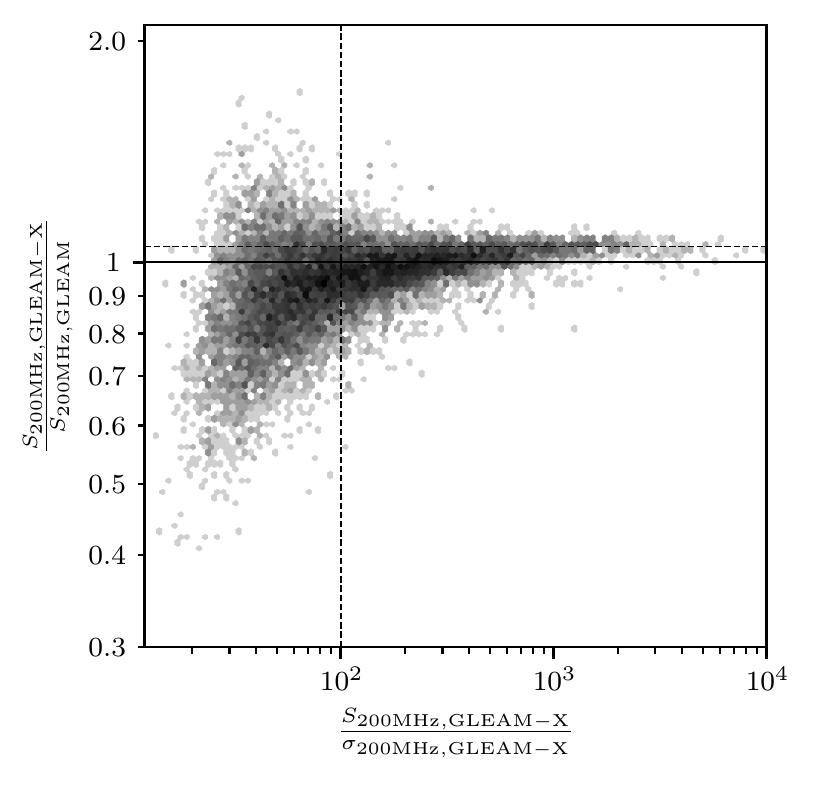}
    \caption{The ratio of the 200-MHz integrated flux densities measured in GLEAM-X and GLEAM, as a function of signal-to-noise in GLEAM-X, for compact sources matched between the two surveys in the region released in this work. 
    The horizontal black line shows a ratio of 1 and the horizontal dashed black line shows a ratio of 1.05, which is a better visual fit at high signal-to-noise.
    The vertical line is plotted at a signal-to-noise of 100, approximately the 90\,\% completeness level of GLEAM in this region.
    %The error bars are the quadrature sum of the fitting errors for each catalogue obtained for each source.
    The error bars are omitted for clarity, but as the quadrature sum of the measurement errors in both surveys, increase to $\sim10$\,\% at the 90\,\% completeness limit of GLEAM, and to $\sim30$\,\% at the faintest levels.}
    \label{fig:GLEAM-X_GLEAM_S200_ratio}
\end{figure}

\Fig~\ref{fig:GLEAM-X_GLEAM_S200_ratio} plots the ratio of the two catalogue integrated flux densities as a function of signal-to-noise in GLEAM-X, with a correction applied to the GLEAM flux densities. The ratio trends toward 1.05 at higher flux densities, although the very brightest sources show only small discrepancies from unity.
 Since the effect is small, we do not attempt to correct for it here, but may revisit our data processing in future to see if it can be reduced, corrected, or eliminated. Since the flux density scale is tied to GLEAM, which has an 8\,\% error relative to other surveys, this value may be used as an error when combining the data with other work.

\begin{figure}
    \centering
    \includegraphics[width=1\linewidth]{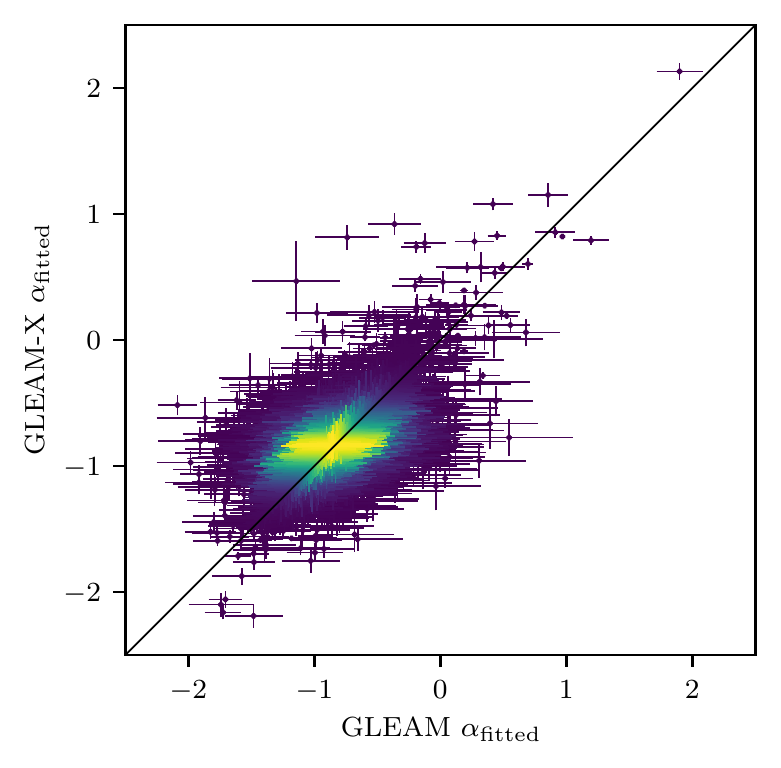}
    \caption{Spectral indices $\alpha$ from the spectra fitted across the 20 7.68-MHz narrow bands of GLEAM-X (ordinate) against GLEAM (abscissa), for compact sources matched between the two surveys in the region released in this work. The color axis shows an arbitrary number density scaling to show where there are more points. The error bars are the fitting errors obtained for each source, dominated by the flux density calibration error at the bright end and the local RMS noise at the faint end. The diagonal line shows a 1:1 ratio.}
    \label{fig:GLEAM-X_GLEAM_alpha}
\end{figure}

No obvious trends are visible in the fitted spectral indices (\Fig~\ref{fig:GLEAM-X_GLEAM_alpha}); we note that the error bars on the GLEAM-X measurements are uniformly smaller due to the increased signal-to-noise of the data.

%\begin{figure}
%    \centering
%    \includegraphics[width=1\linewidth]{figs/GLEA%M-X_GLEAM_S200_comparison.pdf}
%    \caption{200-MHz integrated flux densities fitted across the 20 7.68-MHz narrow bands of GLEAM-X (ordinate) against GLEAM (abscissa), for compact sources matched between the two surveys in the region released in this work. The color axis shows an arbitrary $\log$ number density scaling to show where there are more points. The error bars are the fitting errors obtained for each source, dominated by the flux density calibration error at the bright end and the local RMS noise at the faint end. The diagonal line shows a 1:1 ratio.}
%    \label{fig:GLEAM-X_GLEAM_S200}
%\end{figure}

\subsubsection{Noise properties}\label{sec:noise_properties}

\begin{figure*}
    \centering
    \includegraphics[width=\textwidth]{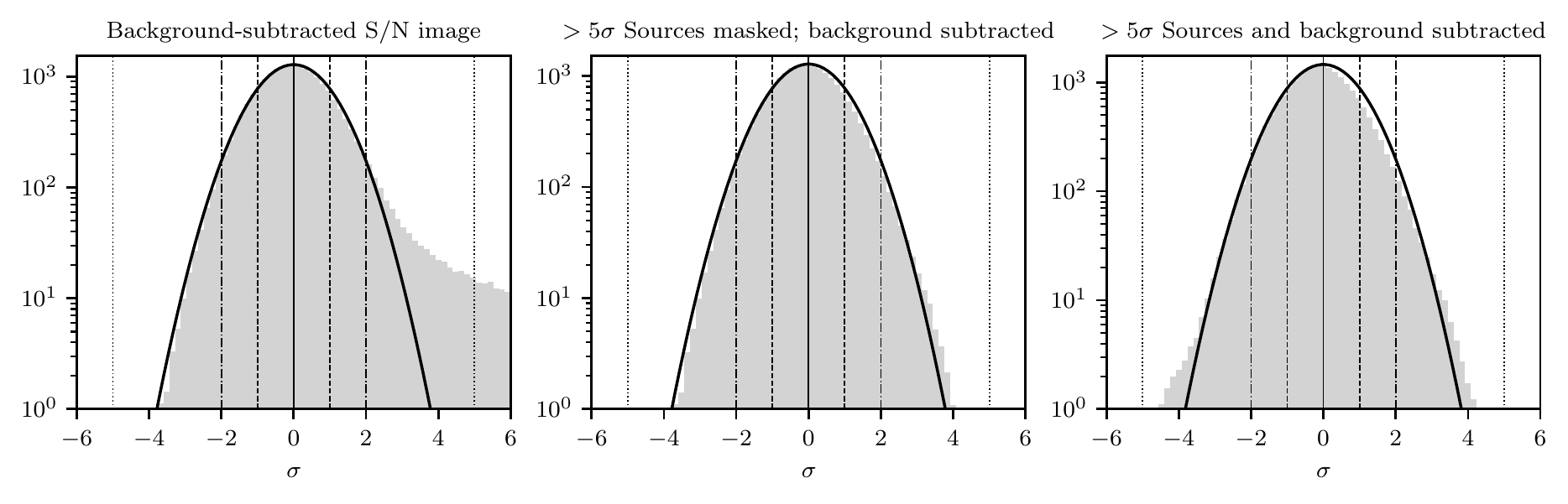}
    \caption{Noise distribution in a typical 18\,square degrees of the wideband source-finding image. \textsc{BANE} measures the average RMS in this region to be 1.06\,mJy\,beam$^{-1}$. To more clearly show any deviation from Gaussian noise, the ordinate is plotted on a log scale. The leftmost panel shows the distribution of the S/Ns of the pixels in the image produced by subtracting the background and dividing by the RMS map measured by \textsc{BANE}; the right panel shows the S/N distribution after masking all sources detected at $5\sigma$ down to $0.2\sigma$. The light grey histograms show the data. The black lines show Gaussians with $\sigma=1$; vertical solid lines indicate the mean values. $|\mathrm{S/N}|=1\sigma$ is shown with dashed lines, $|\mathrm{S/N}|=2\sigma$ is shown with dash-dotted lines, and $|\mathrm{S/N}|=5\sigma$ is shown with dotted lines.}
    \label{fig:noise_distribution}
\end{figure*}

We briefly examine the noise properties of the source-finding 170--231-MHz image. We use a 18\,deg$^2$ region centered on RA~$10^\mathrm{h}30^\mathrm{m}$ Dec$-27^\circ30'$ with fairly typical source distribution. Following \cite{2017MNRAS.464.1146H}, we measure the background of the region using \textsc{BANE}, and subtract it from the image. We then use \textsc{AeRes} (``\textsc{Aegean} REsiduals'') from the \textsc{Aegean} package to mask out all sources which were detected by \textsc{Aegean}, down to $0.2\times$ the local RMS. We also use \textsc{AeRes} to subtract the sources to show the magnitude of the residuals. Histograms of the remaining pixels are shown, for the unmasked and masked images, in \Fig~\ref{fig:noise_distribution}.

The higher resolution of the GLEAM-X survey compared to GLEAM means that confusion forms a smaller fraction of the noise contribution, and thus the noise distribution is almost completely symmetric. Surveys close to the confusion limit will see a skew toward a more positive distribution, as seen by \cite{2017MNRAS.464.1146H}.
Noise and background maps are made available as part of the survey data release.

\subsubsection{Astrometry}\label{sec:overall_astrometry}

Following \cite{2017MNRAS.464.1146H}, we measure the astrometry using the 200-MHz catalogue, as this provides the locations and morphologies of all sources. To determine the astrometry, high signal-to-noise (integrated flux density $>50\sigma$) GLEAM-X sources are cross-matched with the isolated sparse NVSS and SUMSS catalogue (\Sect~\ref{sec:astrometry}); the positions of sources in these catalogues are assumed to be correct and RA and Dec offsets are measured with respect to those positions. The average RA offset is $+14\pm700$\,mas, and the average Dec offset is  $+21\pm687$\,mas (errors are 1~standard deviation).

\begin{figure}
    \centering
    \includegraphics[width=\linewidth]{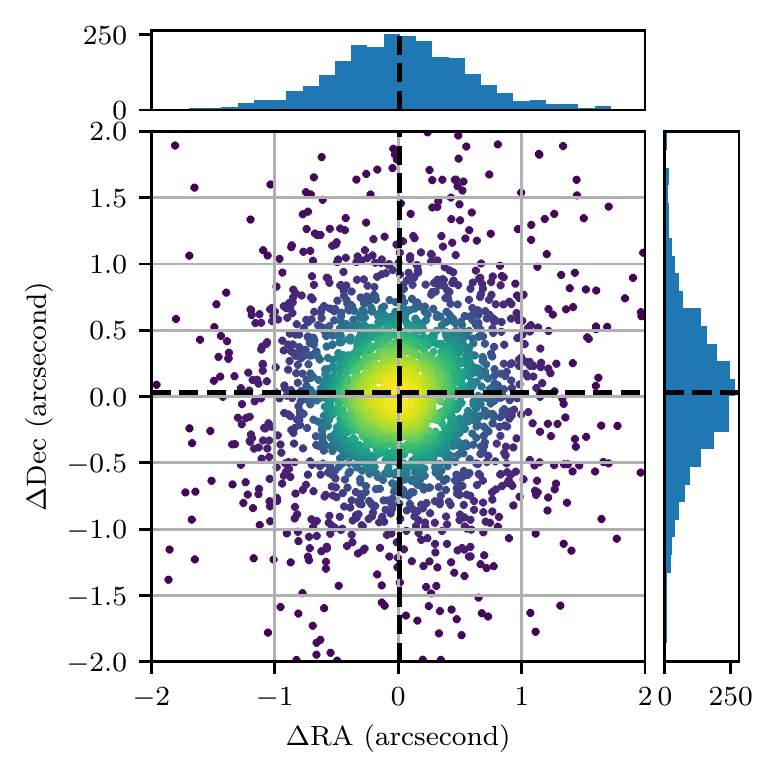}
    \caption{The astrometric offsets of 19,771 isolated, compact, $>50$-$\sigma$ sources after cross-matching against the NVSS and SUMSS reference catalogue described in \sect~\ref{sec:overall_astrometry}. Vertical and horizontal dashed lines indicate the mean offset values in the RA and Dec directions, respectively. Similarly, the horizontal and vertical histograms highlight the counts of the astrometry offsets in each direction.}
    \label{fig:astrometry}
\end{figure}

In 99\,\% of cases, fitting errors on the positions are larger than the measured average astrometric offsets. Given the scatter in the measurements, we do not attempt to make a correction for these offsets. As each snapshot has been corrected, residual errors should not vary on scales smaller than the size of the primary beam. \fig~\ref{fig:astrometry} shows the density distribution of the astrometric offsets, and histograms of the RA and Dec offsets, which were used to calculate the values listed in this section.

\subsection{Completeness and reliability}\label{sec:reliability}

\subsubsection{Completeness}

Following the same procedure as \cite{2017MNRAS.464.1146H}, simulations are used to quantify the completeness of the source catalogue at 200\,MHz, using the wideband mosaics. 
26 realisations are used in which 25,000 simulated point sources of the same flux density were injected into the 170--231\,MHz mosaics (at approximately 20\,\% of the true source density).
The flux density of the simulated sources is different for each realisation, spanning the range $10^{-3}$ to $10^{-0.5}$~Jy in increments of 0.1~dex.
The positions of the simulated sources are chosen randomly but not altered between realisations; to avoid introducing an artificial factor of confusion in the simulations, simulated sources are not permitted to lie within $5'$ of each other. Sources are injected into the mosaics using \textsc{AeRes}. The major and minor axes of the simulated sources are set to $a_\mathrm{psf}$ and $b_\mathrm{psf}$, respectively.

For each realisation, the source-finding procedures described in \Sect~\ref{sec:catalogue} are applied to the mosaics and the fraction of simulated sources recovered is calculated.
In cases where a simulated source is found to lie too close to a real ($>5\sigma$) source to be detected separately, the simulated source is considered to be detected if the recovered source position is closer to the simulated rather than the real source position. This type of completeness simulation therefore accounts for sources that are omitted from the source-finding process through being too close to a brighter source. 

\begin{figure}
\centering
\includegraphics[width=1\linewidth]{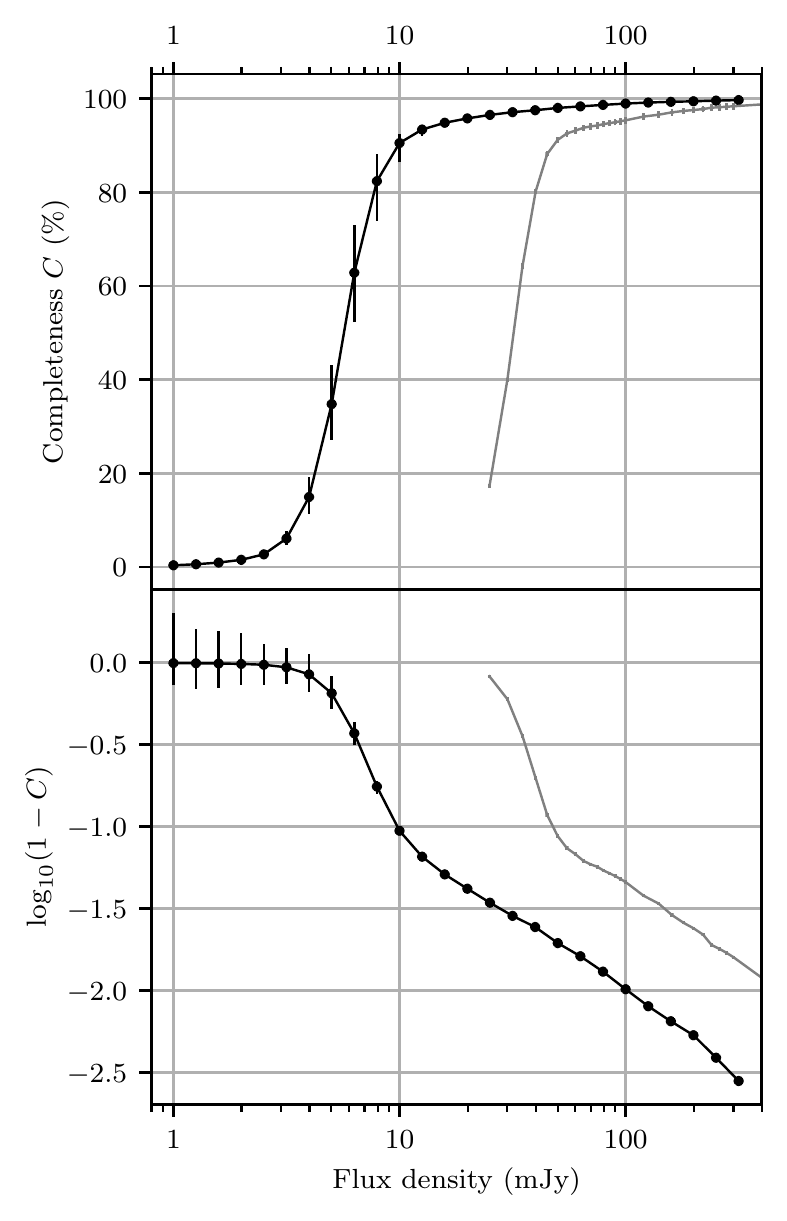}
\caption{Completeness as a function of integrated flux density at 200\,MHz for the region published in this work, which has RMS noise $1.27\pm0.15$\,mJy\,beam$^{-1}$. The top panel shows the completeness $C$; to better display the completeness as it approaches 100\,\%, the bottom panel shows $\log_{10}\left(1-C\right)$. Black markers and lines indicate GLEAM-X; for comparison, grey markers and lines show the completeness of GLEAM for a low-noise region used to determine source counts in GLEAM ExGal ($\sim$2,500\,deg$^2$ with RMS noise $6.8 \pm 1.3$\,mJy\,beam$^{-1}$).}
\label{fig:cmp_vs_s}
\end{figure}

\begin{figure*}
    \centering
    \includegraphics[width=\textwidth]{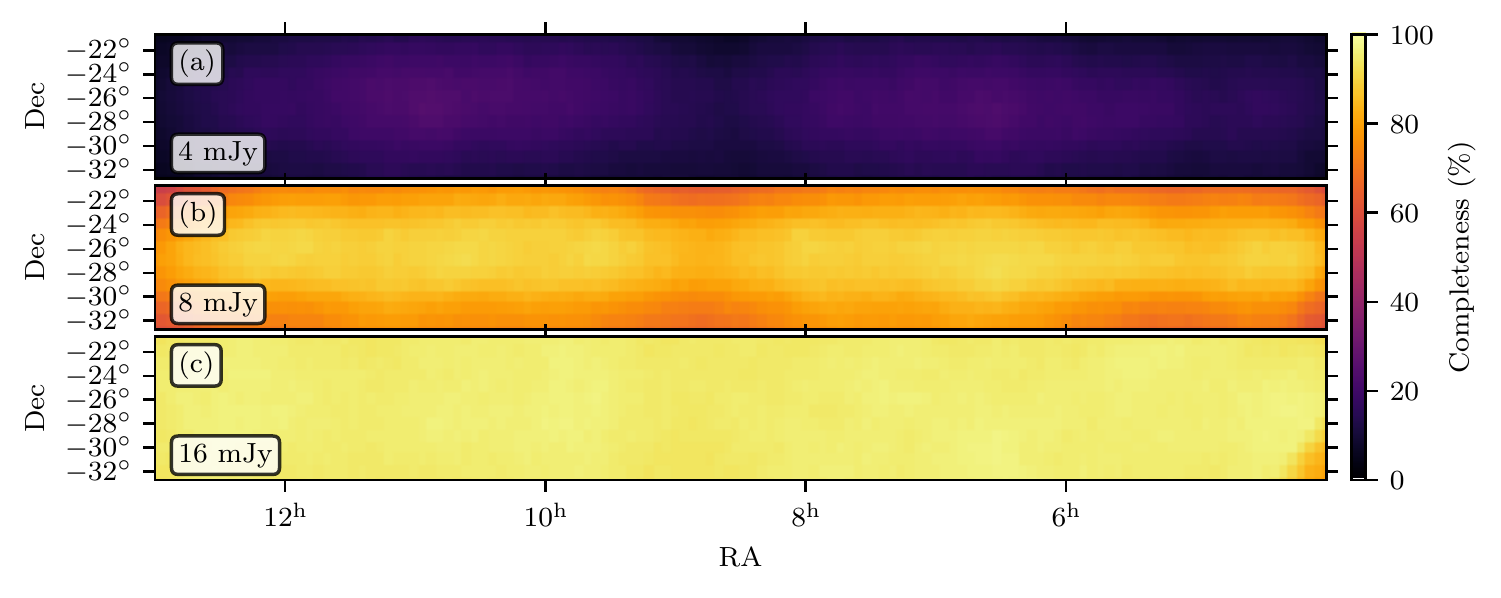}
    \caption{Completeness fraction as a function of position on the sky for three representative cuts in source integrated flux density at 200\,MHz, for the catalogue released in this work.}
    \label{fig:cmp_vs_pos}
\end{figure*}

\fig~\ref{fig:cmp_vs_s} shows the fraction of simulated sources recovered as a function of
$S_{200 \mathrm{MHz}}$. The completeness is estimated to be 50\,\% at $\sim5.6$\,mJy rising to 90\,\% at $\sim10$\,mJy; these flux densities were typically below the RMS noise in GLEAM ExGal. Errors on the completeness estimate are derived assuming Poisson errors on the number of simulated sources detected. \Fig~\ref{fig:cmp_vs_pos} shows the spatial distribution of the completeness for the work presented here; the slight dependence on RA is largely due to the presence of bright sources in large mosaics, e.g. Hydra~A at $\sim$ RA~$09^\mathrm{h}20^\mathrm{m}$ Dec~$-12^\circ$. The roll-off in Declination is due to the primary beam sensitivity of the single drift scan used in this work; in the full survey, multiple drift scans will be used to ensure near-uniform sensitivity and completeness across the sky.

The completeness at any pixel position is given by $C = N_{\mathrm{d}}/N_{\mathrm{s}}$, where $N_{\mathrm{s}}$ is the number of simulated sources in a circle of radius $6\arcdeg$ centred on the pixel and $N_{\mathrm{d}}$ is the number of simulated sources that were detected above $5\sigma$ within this same region of sky.
The completeness maps, in \textsc{fits} format, can be obtained from the supplementary material. Postage stamp images from the GLEAM-X VO server also include the estimated completeness at representative flux densities in their headers.

\subsubsection{Reliability}

\begin{figure}
    \centering
    \includegraphics[width=1\linewidth]{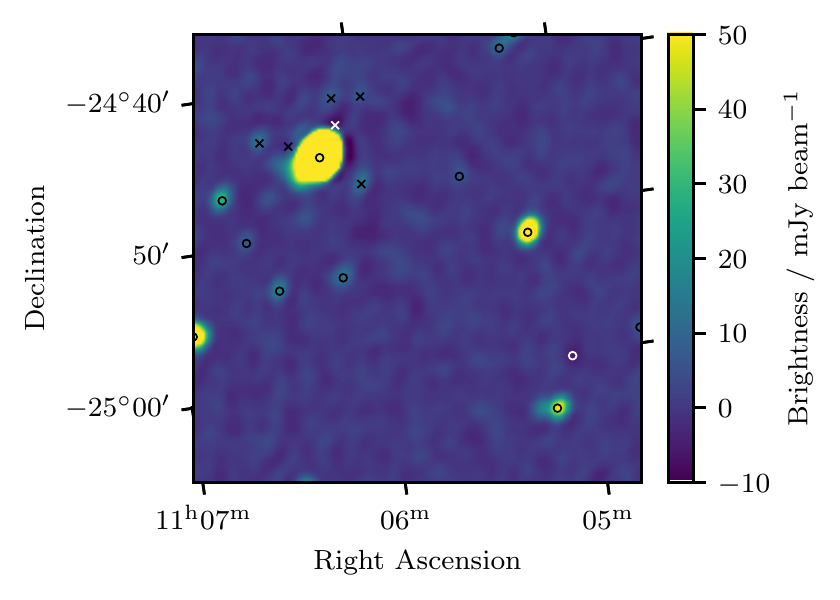}
    \caption{An example of filtering artefacts that present as spurious positive and negative sources around very bright sources. Black circles indicate detected positive components that are not filtered, while black $\times$s show positive components that are filtered. The white circle shows a negative source that was not filtered, while the white $\times$ shows a negative source that was filtered for being too close to a bright source.}
    \label{fig:artefacts}
\end{figure}

To test the reliability of the source finder and check how many of the detected sources might be false detections, we use the same source-finding procedure as described above but search only for negative peaks. \textsc{Aegean} is run with a seedclip of $4\sigma$ (allowing for detections with peaks above this limit) and detections outside of the central region are cut. This initially yields 1,144 negative detections. Filtering the results to retain only sources with integrated flux densities $S_{\mathrm{int}}>5\sigma$ leaves 198~detections. Inspection revealed that some of these detections were artefacts around very bright sources, rather than noise peaks (see \Fig~\ref{fig:artefacts}). There were also similar positive detections of artifacts around these bright sources. We filtered out any detections (positive or negative) that were 
\begin{itemize}
    \item within $5'$ of a positive detection whose peak flux density was $\geq 2\,$Jy and where the absolute value of the ratio of the fainter peak to the bright peak was $\geq 350$; or
   \item within $12'$ of a positive detection whose peak flux density was $\geq 6\,$Jy and where the absolute value of the ratio of the fainter peak to the bright peak was $\geq 650$.
\end{itemize}
This accounts for the moderately bright artefacts closer in to the bright sources and fainter artefacts that can exist further out from very bright sources. This filtering cuts 157~positive detections and 149~negative detections.

\begin{figure}
    \centering
    \includegraphics[width=1\linewidth]{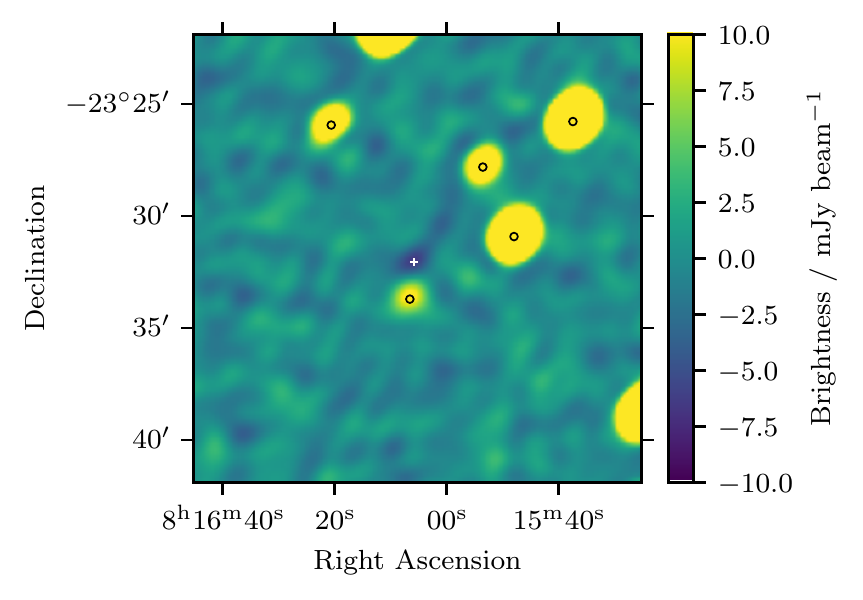}
    \caption{An example of a negative source found next to a positive source that could optionally be filtered when generating the reliability estimate. Black circles indicate detected positive components that are not filtered; the white $+$ shows a negative source that can optionally be filtered.}
    \label{fig:optional_filter}
\end{figure}

We also note that there is a tendency for negative sources to appear close to positive sources regardless of their brightness, potentially due to faint uncleaned sidelobes slightly reducing the map brightness very close to sources. These negative sources will not have positive counterparts, so potentially can also be filtered before estimating the reliability. The criterion in this case is that they cross-match with a positive source within $2'$.
An example is shown in \Fig~\ref{fig:optional_filter}. These comprise a further 46 sources which may optionally be removed.

\begin{figure}
\centering
\includegraphics[width=1\linewidth, angle=0]{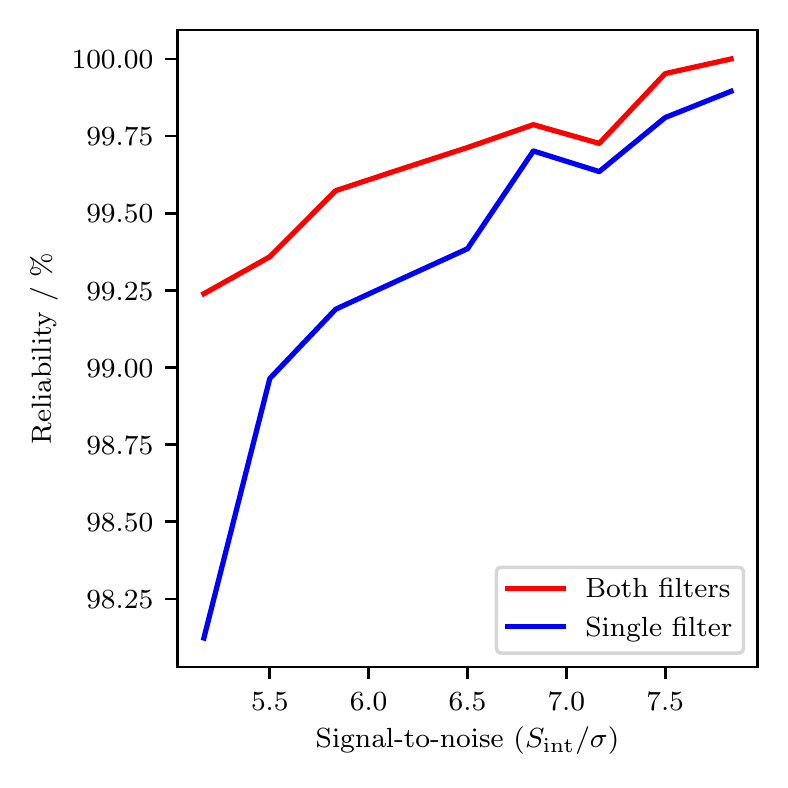}
\caption{Estimates of the reliability of the catalogue as a function of signal to noise. The lower blue curve shows a conservative estimate without filtering negative sources detected on the edges of positive sources. The upper red curve shows a more generous estimate derived after filtering these sources out. In comparison, GLEAM ExGal has a reliability of 98.9\,\%--99.8\,\% at these signal-to-noise levels.}
\label{fig:reliability}
\end{figure}

Comparing the filtered samples of negative to positive detections, we can estimate the number of positive detections that are false detections as a function of signal to noise. For a conservative estimate, where we do not apply the second filter, we find that at a signal-to-noise ratio of five, the number of false detections is just under $2$\,\%, falling quickly to $1$\,\% for $S_{\mathrm{int}}>5.5\sigma$. If we also filter negative sources that lie close to positive sources, we find that the reliability is much higher, with only 0.75\,\% of sources false at 5-$\sigma$, and rising to none at 8-$\sigma$. For each significance bin, we convert these fractions to a reliability estimate and plot them as a function of signal-to-noise in \Fig~\ref{fig:reliability}. We note that were the noise completely Gaussian, we would expect just one $+5\sigma$ source in this sky area to appear purely by chance, and none with flux density $>5.5\sigma$; i.e., a reliability of 99.999\,\% in the faintest bin, rising quickly 100\,\%.

%>>> import numpy as np
%>>> bmaj=77.
%>>> bmin=61.
%>>> bvol = 1.133*(bmaj/3600)*(bmin/3600)
%>>> skyarea = 1447/bvol
% # double-sided statistic (area inside +/- 5-sigma)
%>>> g5 = 0.999999426696856
%>>> skyarea*(1-g5)
%2.0202605623696983
% # double-sided statistic (area inside +/- 5.5-sigma)
%>>> g55 =       0.999999962020875
%>>> skyarea*(1-g55)
%0.13383448032802212

%This confirms that we can be confident to a high degree of certainty that sources with fluxes $\geq 5\sigma$ are genuine. 

%The distribution of negative peaks before and after this filtering can be seen in Fig.~\ref{fig:noisedist}. 
%\begin{figure}
%\centering
%\includegraphics[width=1\linewidth]{figs/GLEAMX_neg.pdf}
%\caption{\textbf{placeholder} Distributions of negative detections in the area published in this work. The blue bars show the distribution before filtering for artefacts and the red bars show is after filtering.} %The black line shows the expected number of detections based on a Gaussian noise distribution.}
%\label{fig:noisedist}
%\end{figure}

%Again using the same procedure as \cite{2017MNRAS.464.1146H}, we use the same source-finding algorithm but invert the brightness, looking only for sources with $S_\mathrm{200MHz}<-5\sigma$. In the \catarea~deg$^2$ region, we find \nneg~negative sources. As the noise distribution is close to symmetric (\Sect~\ref{sec:noise_properties}), we expect to see an approximately equal number of false positive sources in the same area. We thus estimate the catalogue reliability to be:

%$1.0 - \frac{\nneg}{\nsrc} = {\pctreliable}$\,\%

\subsection{Spectral fitting}\label{sec:alpha}

We fit two models to the twenty narrow-band flux density measurements for all detected sources (using $S\propto\nu^\alpha$). The first model is a simple power-law parameterised as 

\begin{equation}
    \label{eq:pl}
    S_\nu = S_{\nu_0}\left(\frac{\nu}{\nu_0}\right)^\alpha,
\end{equation}

\noindent where $S_{\nu_0}$ is the brightness of the source, in Jy, at the reference frequency $\nu_0$, and $\alpha$ describes the gradient of the spectral slope in logarithmic space. We also extend this power-law model to,
\begin{equation}
    \label{eq:cpl}
    S_\nu = S_{\nu_0}\left(\frac{\nu}{\nu_0}\right)^\alpha \exp{\left( q\mathrm{ln}\left(\frac{\nu}{\nu_0}\right)^2\right)},
\end{equation}

\noindent which includes the additional free parameter $q$ to capture any higher order spectral curvature features, where increasing $|q|$ captures stronger deviations from a simple power law; if q is positive, the curve is opening upward (convex) and if q is negative, the curve is opening downward (concave). This model is not physically motivated, and may not appropriately describe sources with different power-law slopes in the optically thin and thick regimes, but provides a useful filter to identify interesting sources. For both models we set $\nu_0$ to 200-MHz. 

To perform accurate spectral fitting, the errors on the flux density measurements must be known. Following \cite{2017MNRAS.464.1146H}, spectral fitting allows us to check the flux density consistency of the catalogue. A flux density scaling error of 2\,\% yields a median reduced $\chi^2$ of unity across the catalogue, whereas higher or lower values bias the reduced $\chi^2$ lower or higher as a function of signal-to-noise. We thus adopt 2\,\% as the measure of our internal flux density scale, and set the errors on the flux density to this value added in quadrature with the local fitting error from \textsc{Aegean}. (Note that 8\,\% is more appropriate when comparing with other catalogues as this is the flux density scale accuracy of GLEAM, to which GLEAM-X is tied (see \Sect~\ref{sec:gleam}).)

% The spectral peak, $\nu_p$, relates to the curvature parameter $q$ as $\nu_p = \mathrm{exp}^{-\alpha/2q}$

\begin{figure*}
    \centering
   \includegraphics[width=\textwidth]{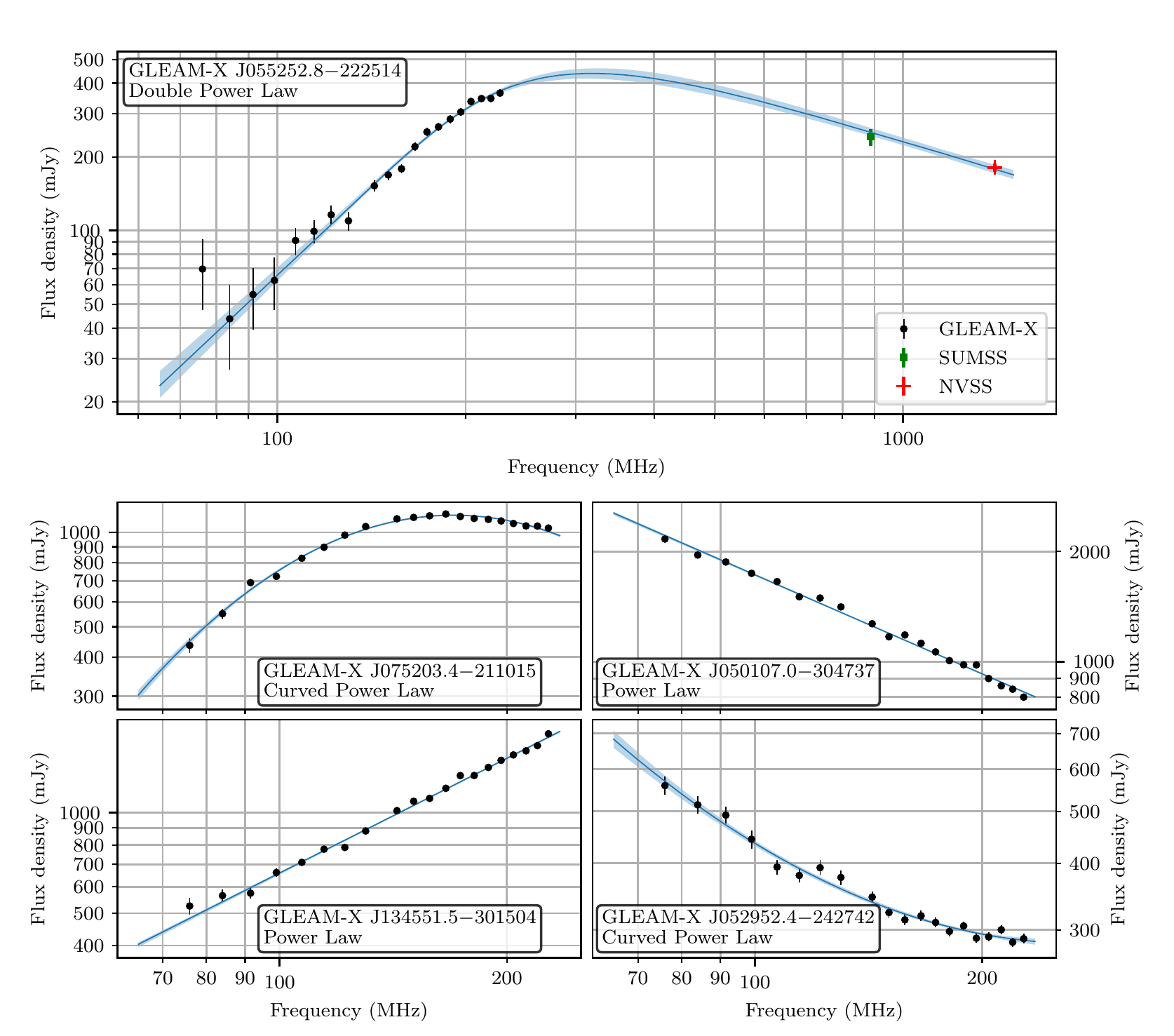}
    \caption{Five example SEDs of sources selected to highlight the variety of spectral shapes seen within the GLEAM-X data. The panel inset includes the source name and model used to fit the presented data. The optimised model and its 1-$\sigma$ confidence interval is overlaid as the blue line of each source. The `Power Law' and `Curved Power Law' are defined as \Eqns~\ref{eq:pl} and \ref{eq:cpl}, respectively. The `Double Power Law' used for GLEAM-X\,J$055252.8-222514$ shows \Eqn~3 of \citet{2017ApJ...836..174C} fitted using the GLEAM-X data and higher frequency measurements from SUMSS and NVSS.  
    \label{fig:example_SEDs}}
\end{figure*}

We applied the Levenberg-Marquardt non-linear least-squares regression algorithm  \citep[as implemented in the \textsc{scipy} \textsc{python} module;][]{2020SciPy-NMeth} to \Eqns~\ref{eq:pl} and \ref{eq:cpl} for each detected source. We did not include narrow-bands with negative integrated flux density measurements. We discarded the fitting results if
\begin{itemize}
\item there were fewer than 15 integrated flux density measurements for a source;
\item a $\chi^2$ goodness-of-fit test indicated at a $>99$\,\% likelihood of an incorrectly-fit model; or
\item $q/\Delta q<3$, to ensure constrained deviations from a power-law are statistically significant.
\end{itemize}
For this initial data release we included only the model with the lower reduced-$\chi^2$ statistic in our catalogue. Applying these criteria a total of \nplfit{} and \ncplfit{} source components have fitting results recorded for power-law and curved power-law models, respectively. \Fig~\ref{fig:example_SEDs} shows five example  SEDs, four with either power-law or curved power-law models constrained using exclusively GLEAM-X, and one with GLEAM-X data supplemented with data from SUMSS and NVSS to fit a two-component power-law model described as

% \begin{equation}
\begin{align*}
    \label{eq:2pl}
    S_\nu = \frac{S_p}{1-\exp{\left(-1\right)}}\times\\\left(1 - \exp{\left(-\left(\frac{\nu}{\nu_p}\right)^{\alpha_{\mathrm{thin}}-\alpha_{\mathrm{thick}} }\right)}\right)    \left(\frac{\nu}{\nu_p}\right)^{\alpha_{\mathrm{thick}}}
\end{align*}
% \end{equation}

\noindent where $S_p$ is the brightness (Jy) at the peak frequency $\nu_p$ (MHz), and $\alpha_{\mathrm{thin}}$ and $\alpha_{\mathrm{thick}}$ are the spectral slopes in the optically thin and optically thick regimes, respectively \citep{2017ApJ...836..174C}.

\begin{figure}
    \centering
   \includegraphics[width=1\linewidth]{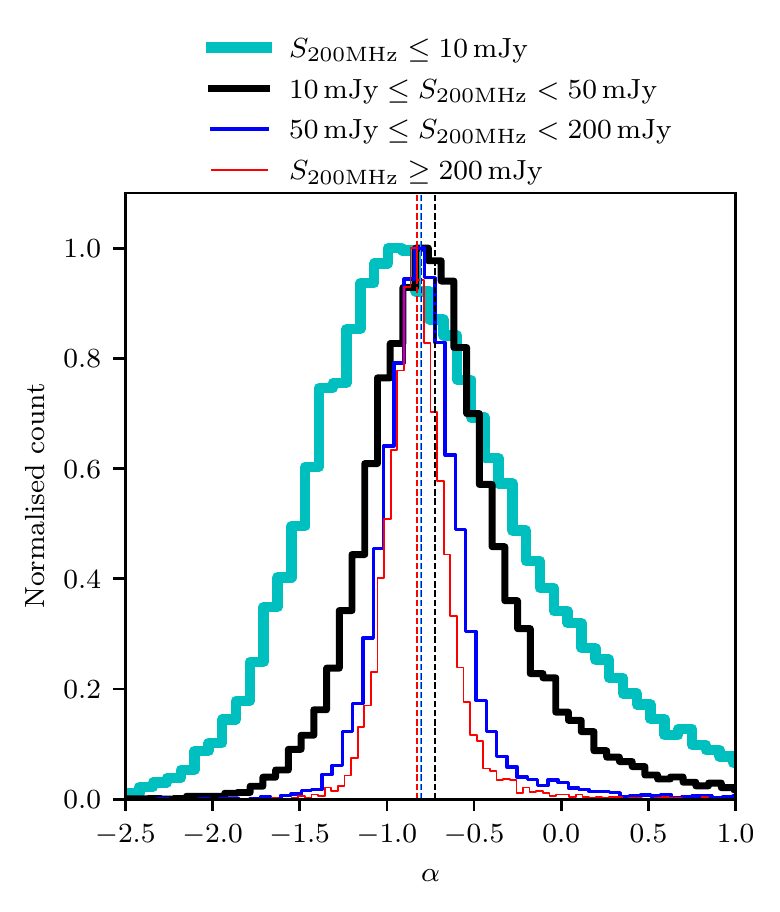}
    \caption{The spectral index distribution calculated for sources where the fit was successful (reduced~$\chi^2<1.93$). The cyan line shows sources with $S_\mathrm{200MHz}<10$\,mJy, the
black line shows sources with $10\leq S_\mathrm{200MHz}<50$\,mJy, the blue line shows sources with $50\leq S_\mathrm{200MHz}<200$\,mJy, and the red line shows sources with $S_\mathrm{200MHz}>200$\,mJy. The dashed vertical lines of the same colours show the median values for each flux density cut: $-0.81$, $-0.73$, $-0.80$, and $-0.83$, respectively.
    \label{fig:alpha_distribution}}
\end{figure}

For sources fit well by power-law SEDs, the distributions of spectral indices $\alpha$ with respect to flux density are plotted in in \Fig~\ref{fig:alpha_distribution}. The median $\alpha$ for the brightest bin is $-0.83$, in excellent agreement with previous results \citep[e.g.][]{2003MNRAS.342.1117M,2014MNRAS.440..327L,2015A&A...582A.123H}.

\begin{figure}
    \centering
    \includegraphics[width=\linewidth]{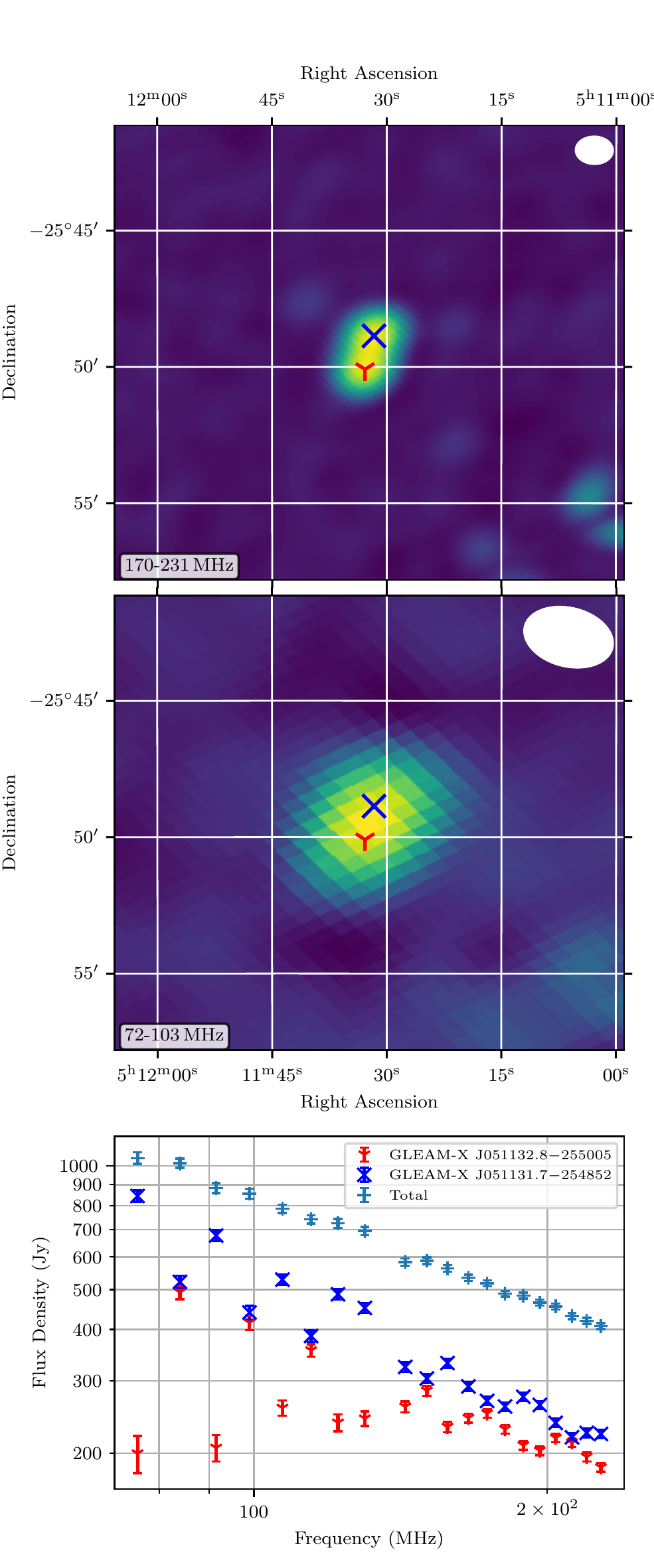}
    \caption{An example where the \textsc{aegean} priorised fitting routine was unable to produce consistent flux density measurements across sub-bands for a pair of components (GLEAM-X\,J$051132.8-255005$ and GLEAM-X\,J$051131.7-254852$) belonging to a single island. The top and middle panels are the 170-231\,MHz and 72-103\,MHz images towards these components, respectively, and the white ellipse in the upper right corner is the corresponding PSF. The bottom panel contains the SEDs of the individual components and their total.  }
    \label{fig:bad_sed}
\end{figure}

The priorised fitting routine in \textsc{aegean} separates the island finding stage from the component characterisation stage, and is analogous to aperture photometry in optical images \citep{2018PASA...35...11H}.  We use this in GLEAM-X to ensure that each radio-component identified in our deep 170--231\,MHz source finding image has an equivalent component characterisation in each of the other 25~GLEAM-X images. This process however does not enforce spectral smoothness between images adjacent in frequency. For GLEAM-X, this process becomes less reliable towards lower frequencies, where the PSF becomes large enough that nearby components are blended to the point where their brightness profiles can not be distinguished. Although model optimisation methods may be able to constrain the total brightness across all components, the brightness between individual components become degenerate. We highlight an example of this behaviour in Figure~\ref{fig:bad_sed}.  This problem is most apparent for sources that are slightly resolved and characterised as two separate components within 120$''$ from one another. Further development of \textsc{aegean} to perform component characterisation across all images jointly while including physically-motivated parametisation of the spectra is planned to address this issue. 

\subsection{Final catalogue}

The resulting catalogue consists of \nsrc~radio sources detected over \survarea~deg$^2$. \nfit{}~sources are fit well by power-law or curved-spectrum SEDs. The catalogue has \ncol{}~columns (see Appendix~\ref{sec:columns}) and is available via Vizier. The catalogue measurements can be used to perform more complex spectral fits, especially in conjunction with other radio measurements. \Tab~\ref{tab:survey_stats} shows the properties of the images and catalogue in this data release, as well as some forward predictions for the full survey, in comparison to GLEAM.

% Nights in IDR1   RA start    RA end
% 2018-02-04         50           220
% 2018-02-09         40           215
% 2018-02-20         60           240
% 2018-03-05         85           270
% So at maximum 50 to 240
% Also need to subtract ~10 deg from each end because of PB cutoff
% 60 to 215
% Visually, image quality is much lower past 190, probably because of Cen A
% So let's go with 60 to 195 = 4h to 13h
% +/- 6deg highlights the noise levels we can hope for
% TopCat: ra >= 4*15 && ra <= 13*15 && dec <= -20.7 && dec >= -32.7
% TopCat: ref_ra >= 4*15 && ref_ra <= 13*15 && ref_dec <= -20.7 && ref_dec >= -32.7

% Total survey area: 20636*(1+np.sin(np.radians(30.)))

\begin{table*}
\centering
    \caption{Survey properties and statistics, for the region published in this paper, in comparison to the largest single data release from GLEAM \citep{2017MNRAS.464.1146H}, and estimates for the full survey. Values are given as the mean, $\pm$ the standard deviation where appropriate. The statistics shown are derived from the wideband (170--231\,MHz) image. The internal flux density scale error applies to all frequencies. \label{tab:survey_stats}}
    \begin{tabular}{cccc}
    \hline
    Property & GLEAM-X (this release) & GLEAM ExGal & GLEAM-X (full) \\
    \hline
    Number of sources & \nsrc{} & 307,456 & 1.7\,M \\
    Number of sources spectrally fit & \nfit{} & 254,453 & 1.5\,M \\
    Sky area & \survarea{}~deg$^{2}$ & 24,402~deg$^{2}$ & 30,954~deg$^{2}$ \\
    Source density & 55~deg$^{-2}$ & 13~deg$^{-2}$ & 55~deg$^{-2}$ \\
    RA astrometric offset & $+14\pm700$\,mas & $-4\pm16''$ & $\sim20\pm700$\,mas\\
    Dec astrometric offset  & $+21\pm687$\,mas & $0.1\pm3.6''$ & $\sim20\pm700$\,mas \\
    Internal flux density scale error & 2\,\% & 2\,\% & 2\,\% \\
    50\,\% completeness & 5.6\,mJy & 55\,mJy & 5.6\,mJy \\
    90\,\% completeness & 10\,mJy & 170\,mJy & 10\,mJy  \\ 
    98\,\% completeness & 50\,mJy & 500\,mJy  & 50\,mJy \\ 
    Reliability for $S_\mathrm{int}\geq7\sigma$ & \pctreliablehigh{}\,\% & 99.8\,\%  & \pctreliablehigh{}\,\% \\
    Reliability for $S_\mathrm{int}\geq5\sigma$ & \pctreliablelow{}\,\% & 98.9\,\% &  \pctreliablelow{}\,\% \\
    Image RMS noise & $1.27\pm0.15$\,mJy\,beam$^{-1}$ & $11.3\pm7.3$\,mJy\,beam$^{-1}$  & $\sim1.2$\,mJy\,beam$^{-1}$  \\
    PSF major axis & $77\pm12''$ & $152\pm25''$ & $\sim75$--$110''$ \\
    PSF minor axis & $61\pm6''$ & $134\pm12''$ & $\sim60''$ \\
    \hline
    \end{tabular}
\end{table*}

\section{Extensions to continuum processing}\label{sec:extensions}

The total data volume of GLEAM-X visibilities is large ($\sim2$\,PB) and file transfer operations comprise a significant proportion ($\sim40$\,\%) of our processing time. When processing the data, each observation takes up $\sim100$\,GB of disk space in visibilities, images, and metadata. Given the richness of the GLEAM-X survey, we are strongly motivated to perform additional operations on the data while they reside on disk in order to avoid moving the data more frequently. In this section we discuss the current extensions to the pipeline that we expect will yield a range of science outcomes not possible with mosaicked images.

\subsection{Transient imaging}\label{sec:transients}

The wide field-of-view of the MWA combined with the repeated drift scanning strategy of GLEAM-X yields a dataset that is interesting to search for transient radio sources. \cite{2017PASA...34...20M} compared the first GLEAM catalogue with TGSS-ADR1 and found a single transient candidate, but understanding its nature was difficult with the (limited) data available. Historically this has been a common occurrence for low-frequency radio transients, with many unusual phenomena detected but never fully understood \citep[e.g.][]{2005Natur.434...50H,2016MNRAS.456.2321S,2019ApJ...874..151V}.

The GLEAM-X drift scans were observed such that the LST was matched for repeated observations at the same pointing and frequency. This enabled a search using ``visibility differencing'', wherein calibrated measurement sets were differenced, and the resulting nearly-empty visibilities were inverted to form a dirty image, which could be used to search for transient sources (Honours thesis: O'Doherty 2022; Hancock et al. in prep.). One high-significance candidate was followed up using the large MWA archive, resulting in the discovery of a new type of highly polarised radio transient, repeating on the unusual timescale of 18.18\,minutes \citep{2022Natur.601..526H}. The wide bandwidth of GLEAM-X was key to finding the dispersion measure of the source, and therefore estimating its distance.

The visibility differencing approach resulted in a large number of false positives due to the differences in ionospheric conditions between observations. The discovery of a new type of radio transient, and the utility of our polarisation and wideband measurements, motivates the inclusion of a transient imaging step in our routine pipeline processing.

Our approach is to image every 4-s interval of each observation, at the same time subtracting the deep model that was formed during imaging (\Sect~\ref{sec:images}), the same approach that is currently used for imaging MWA interplanetary scintillation observations (Morgan et al. in prep.). This results in a thermal-noise-dominated Stokes-I image cube where only differences between each time step and the continuum average are recorded. This cube is then stored in an HDF5 file\footnote{\url{http://www.hdfgroup.org/HDF5}} as described in Appendix~2 of \citet{2018MNRAS.473.2965M}. Briefly, the image cube is reordered so that time is the fastest axis, and the pixel data is demoted to half precision (16-bit) floats. This results in a typical data volume of 600\,MB per observation. Once in this format, any number of algorithms can be conveniently applied to detect and measure time-domain signals.  

While imaging every 0.5-s sample would be ideal, it would multiply by $8\times$ the storage and processing requirements for all other steps of the pipeline, but if a signal of interest is discovered then it is simple (and indeed necessary) to reprocess the data with higher time (and, if needed, frequency) resolution. Future data releases will provide these data and quantitative analyses thereof.

\subsection{Binocular imaging}\label{sec:binocular}

The source position offsets determined during the de-warping process (\Sect~\ref{sec:astrometry}) yield information about the slant total electron content (dTEC) averaged over the telescope array projected on to the sky in that field-of-view. If dTEC varies significantly over the array, the wavefronts from different parts of the sky will arrive at different times, and radio sources will appear stretched, duplicated, or will disappear completely. Conversely, if images are created using sub-arrays of the telescope, the apparent difference in source positions can be used to constrain an approximate height of the distorting screen \citep{2015GeoRL..42.3707L,2020RaSc...5507106H}. We thus add a module to the imaging pipeline to routinely produce these binocular images.

\begin{figure}
    \centering
    \includegraphics[width=1\linewidth]{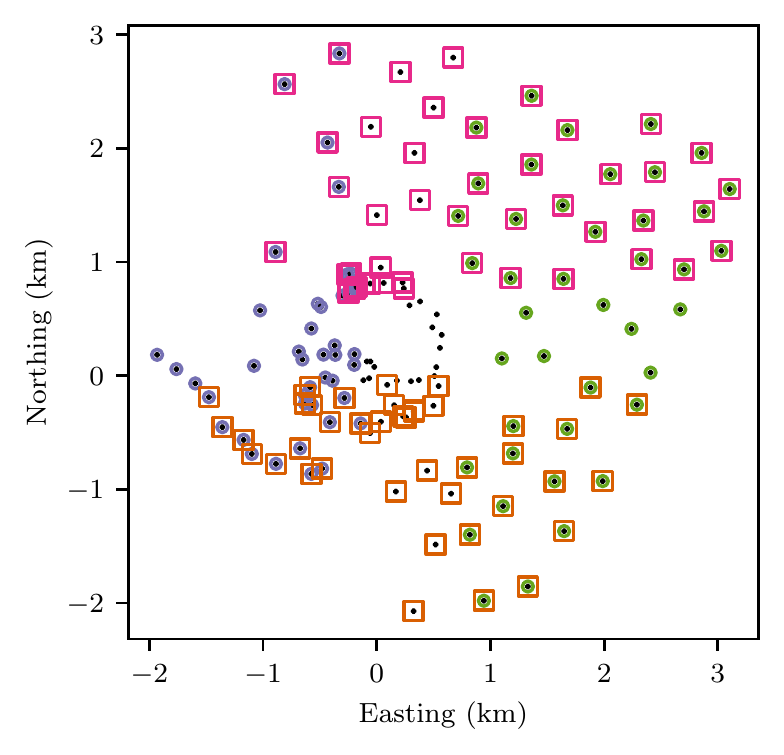}
    \caption{The layout of the tiles comprising the MWA Phase~II extended configuration, overlaid with symbols representing sub-arrays used for ionospheric binocular imaging (\Sect~\ref{sec:binocular}). Pink and orange squares show the North and South pair, while green and lavender circles indicate the East and West pair.}
    \label{fig:binocular_subarrays}
\end{figure}

In choosing the sub-arrays from the extended Phase~\textsc{II}, we face a compromise between sensitivity (higher for large sub-arrays) and parallax lever-arm (better for widely-separated sub-arrays). Additionally we have no prior knowledge of what ionospheric activity will be observed on the night, nor the resources to adjust the imaging to match at the time of processing. To form a generally useful product, we split the array into two pairs of sub-arrays following the cardinal directions, shown in \Fig~\ref{fig:binocular_subarrays}. Each group of 43 or 44~antennas is imaged separately, and source-finding is performed using the default settings of \textsc{Aegean}. These catalogues can form a useful input to future analyses of the ionosphere above the Murchison Radioastronomy Observatory; the data and analysis will be released in future work.

\section{Outlook and conclusions}\label{sec:conclusions}

In this work we described GLEAM-X, a new wideband low-frequency all-southern-sky survey performed using the MWA, as well as the data reduction steps we expect to use to produce a range of continuum data products over 72--231\,MHz. Polarisation data will be described in the upcoming paper by Zhang et al. (in prep). Extensions to our data reduction pipeline to perform transient searches (\sect~\ref{sec:transients}) and binocular imaging (\sect~\ref{sec:binocular}), as well as joint deconvolution of the Galactic Plane (\Fig~\ref{fig:Vela}) will further enhance the capabilities of the survey.

To demonstrate the quality and attributes of the images and catalogues that will be produced by GLEAM-X, we release here \survarea{}~deg$^{2}$ of sky in the form of 26~mosaics across 72--231\,MHz of bandwidths 60, 30, and 8\,MHz, with RMS noises ranging from 15 to just over 1\,mJy\,beam$^{-1}$. Additionally, we form a catalogue of \nsrc{}~sources, \nplfit{} of which are well-fit across our band with power-law spectral energy distributions, and \ncplfit{} with curved power-law spectra. Extrapolating our source density of \srcdensity{}\,deg$^{-2}$ to the $\sim$31,000~deg$^{2}$ that GLEAM-X will eventually cover, we expect to detect of order 1.7\,M sources, and produce $\sim1.5$\,M radio spectra.

We plan to release the survey in a series of data releases; the next will comprise a large ($\sim15$,000 deg$^{2}$) set of images and catalogues covering the southern extragalactic sky centered on the South Galactic Pole (Galvin et al. in prep); secondly we aim to process and release the complete Galactic Plane (Hurley-Walker et al. in prep); finally, we will aim to produce contiguous all-sky coverage. Polarisation, transient, and ionospheric data releases and analyses will also proceed over coming years.

These data will enable a range of science outcomes, some of which are outlined by \cite{2019PASA...36...50B} in their review of scientific opportunities with Phase~\textsc{II} of the MWA. For instance, there is strong potential to detect $10^4$ peaked-spectrum sources in GLEAM-X data, an order of magnitude more than discovered by GLEAM \citep{2017ApJ...836..174C}, and also probing a population an order of magnitude fainter. Improved signal-to-noise on sources with curved and peaked spectra can provide more efficient selection of high-redshift radio galaxies \citep{2020PASA...37...26D}. Many local star-forming galaxies will be resolved, enabling better understanding of the interplay between thermal and non-thermal processes in their energy budgets \citep{2017ApJ...838...68K,2018MNRAS.474..779G}.

The extended configuration of the Phase~\textsc{II} MWA has already been used very capably for targeted  investigations of the extragalactic sky, such as determining the remnant radio galaxy fraction in one of the Galaxy and Mass Assembly fields \citep{2021PASA...38....8Q} and detecting diffuse non-thermal emission in galaxy clusters \citep{2021PASA...38...53D}. Similar studies over the whole sky, particularly exploiting synergies with other recent wide-area surveys such as RACS, are likely to be highly productive. The higher source density of GLEAM-X will for the first time enable cosmological measurements with the MWA. We can resolve the tension between the angular clustering observed with NVSS and TGSS-ADR1 \citep{2019A&A...623A.148D}, investigate differential source counts \citep{2015PhRvD..91d3507C}, and by cross-correlating with measurements of the Cosmic Microwave Background, search for the effects of dark energy via the integrated Sachs-Wolfe effect \citep{1967ApJ...147...73S}. Additionally, GLEAM-X may help to improve sky models for studies of the Epoch of Reionisation, by measuring source brightnesses below 100\,MHz, imaging slightly deeper, and separating sources into more components than LoBES \citep{2021PASA...38...57L}.

Continuum Galactic science shows promise with MWA Phase~\textsc{II} \citep{2022MNRAS.510..593T}, and given the excellent results from our initial exploration of jointly deconvolving GLEAM and GLEAM-X, we expect to make new detections of supernova remnants \citep[SNRs; see e.g.][]{2019PASA...36...45H} and improve measurements of cosmic ray electrons in the Galactic Plane \citep[following][]{2018MNRAS.479.4041S}. Additionally the improved resolution, sensitivity, and wide bandwidth will make possible the examination of the unshocked ejecta of SNRs \citep{2018A&A...612A.110A} and interactions with their environments \citep{2021A&A...653A..62C} via measurements of low-frequency thermal absorption. This creates excellent synergy with TeV observations by the High Energy Stereoscopic System \citep{2004NewAR..48..331H,2006ApJ...636..777A} and the upcoming Cherenkov Telescope Array \citep{2013APh....43....3A} to search for sites of cosmic ray acceleration in our Galaxy \citep[e.g.][]{2019ApJ...885..129M}.

The repeated, overlapping epochs of GLEAM-X and its drift scan observing strategy make it possible to explore radio transients and variability on timescales from seconds to years; comparisons to GLEAM enable a seven-year lever arm. Combining these cadences with the wide bandwidth will enable improved investigation of the startling variability of peaked-spectrum sources found by \cite{2021MNRAS.501.6139R}, and enable distance measurements for dispersion-smeared pulsed transients \citep{2022Natur.601..526H}. As evinced by the latter work, GLEAM-X opens new parameter space in the low-frequency radio sky, and potentially enables further serendipitous discoveries beyond our ability to predict.

\begin{acknowledgements}

We thank the anonymous referee for their comments, which improved the quality of this paper.
NHW is supported by an Australian Research Council Future Fellowship (project number FT190100231) funded by the Australian Government. KR acknowledges a Doctoral Scholarship and an Australian Government Research Training Programme scholarship administered through Curtin University. DK was supported by NSF grant AST-1816492. CJR acknowledges financial support from the ERC Starting Grant `DRANOEL', number 714245. This scientific work makes use of the Murchison Radio-astronomy Observatory, operated by CSIRO. We acknowledge the Wajarri Yamatji people as the traditional owners of the Observatory site. Support for the operation of the MWA is provided by the Australian Government (NCRIS), under a contract to Curtin University administered by Astronomy Australia Limited. Establishment of the Murchison Radio-astronomy Observatory and the Pawsey Supercomputing Centre are initiatives of the Australian Government, with support from the Government of Western Australia and the Science and Industry Endowment Fund. We acknowledge the Pawsey Supercomputing Centre which is supported by the Western Australian and Australian Governments and the China SKA Regional Center prototype at Shanghai Astronomical Observatory which is funded by the Ministry of Science and Technology of China (under grant number 2018YFA0404603) and Chinese Academy of Sciences (under grant number 114231KYSB20170003). Access to Pawsey Data Storage Services is governed by a Data Storage and Management Policy (DSMP). ASVO has received funding from the Australian Commonwealth Government through the National eResearch Collaboration Tools and Resources (NeCTAR) Project, the Australian National Data Service (ANDS), and the National Collaborative Research Infrastructure Strategy. This paper makes use of services or code that have been provided by AAO Data Central (datacentral.org.au). This research has made use of NASA’s Astrophysics Data System Bibliographic Services. The following software was used in this work:
{\sc aoflagger} and {\sc cotter} \citep{2012A+A...539A..95O}; \textsc{WSClean} \citep{2014MNRAS.444..606O,2017MNRAS.471..301O}; {\sc Aegean} \citep{2018PASA...35...11H}; {\sc miriad} \citep{Miriad}; {\sc TopCat} \citep{Topcat} \textsc{NumPy} \citep{NumPy,harris2020array}; \textsc{AstroPy} \citep{Astropy}; \textsc{SciPy} \citep{SciPy}, \textsc{Matplotlib} \citep{Matplotlib}. This work was compiled in the very useful online \LaTeX{} editor Overleaf.

\end{acknowledgements}

\begin{appendix}

% https://tex.stackexchange.com/questions/118606/numbering-tables-a1-a2-etc-in-latex
\setcounter{table}{0}
\renewcommand{\thetable}{A\arabic{table}}

\section{OBSERVATIONS}
\topcaption{GLEAM-X observing summary. The HA and Dec are fixed to the locations shown and the sky drifts past for the observing time shown. Observations typically start just after sunset and stop just before sunrise. The four nights published in this work are shown in bold font. Nights identified as having high ionospheric activity are marked with a ``*''. \label{tab:obs}}
\tablefirsthead{\toprule Date & HA & Dec ($\arcdeg$) & Observing time (hours) \\ \midrule}
\tablehead{\multicolumn{4}{c}{{\bfseries  Continued from previous column}} \\ \toprule 
Date & HA & Dec ($\arcdeg$) & Observing time (hours) \\ \midrule}
\tabletail{ \midrule \multicolumn{4}{c}{{Continued on next column}} \\ \midrule}
\tablelasttail{\bottomrule}
\begin{supertabular}{cccc}
%\caption{GLEAM-X observing summary.
%$N_\mathrm{flag}$ is the number of flagged tiles out of the 128~available.
%\label{tab:obs}}
%\begin{tabular}
%\hline
%Date & HA &  Dec ($\arcdeg$) & Observing time (hours)\tabularnewline
%RA range (h) &
%\hline
%\hline
2018-01-26 & 0 & $+20$ & 8.4 \\
2018-01-27* & 0 & $+1$ & 8.6 \\
2018-01-28* & 0 & $-12$ & 8.7 \\
2018-02-01* & 0 & $+20$ & 8.8 \\
2018-02-02* & 0 & $+1$ & 7.3 \\
2018-02-03* & 0 & $-12$ & 5.8 \\
\textbf{2018-02-04} & 0 & $-26$ & 8.7\\
2018-02-05 & 0 & $-40$ & 3.2 \\
2018-02-06 & 0 & $-55$ & 3.6 \\
2018-02-07 & 0 & $-71$ & 5.7 \\
\textbf{2018-02-09} & $+1$ & $-26$ & 8.6 \\
2018-02-10 & $+1$ & $-12$ & 8.1 \\
2018-02-15* & $+1$ & $+1$ & 9.0 \\
2018-02-16 & $+1$ & $-40$ & 8.9 \\
2018-02-17 & $+1$ & $+20$ & 7.4 \\
2018-02-18 & $+1$ & $-55$ & 8.9 \\
2018-02-19 & $+1$ & $-71$ & 9.0 \\
\textbf{2018-02-20} & 0 & $-26$ & 9.0 \\
2018-02-22 & 0 & $-12$ & 8.9 \\
2018-02-28 & 0 & $-71$ & 9.1 \\
2018-03-01 & 0 & $+1$ & 9.1 \\
2018-03-02 & 0 & $-40$ & 8.8 \\
2018-03-03 & 0 & $+20$ & 9.0 \\
2018-03-04 & 0 & $-55$ & 9.1 \\
\textbf{2018-03-05} & $-1$ & $-26$ & 9.1 \\
2018-03-07 & $-1$ & $-12$ & 9.3 \\
2018-03-10 & $-1$ & $+1$ & 9.5 \\
2018-03-11 & $-1$ & $-40$ & 9.5 \\
2018-03-12 & $-1$ & $+20$ & 9.5 \\
2018-03-13 & $-1$ & $-55$ & 9.6 \\
2018-03-16* & $-1$ & $-71$ & 9.7 \\
2018-05-03 & 0 & $-26$ & 10.7 \\
2018-05-04 & 0 & $-12$ & 10.6 \\
2018-05-05 & 0 & $+1$ & 10.8 \\
2018-05-06 & 0 & $-40$ & 10.9 \\
2018-05-07 & 0 & $+20$ & 10.9 \\
2018-05-08 & 0 & $-55$ & 10.7 \\
2018-05-09 & 0 & $-71$ & 10.7 \\
2018-05-10 & $-1$ & $-26$ & 10.7 \\
2018-05-11 & $-1$ & $-12$ & 10.8 \\
2018-05-12* & $-1$ & $+1$ & 10.7 \\
2018-05-13 & $-1$ & $-40$ & 10.7 \\
2018-05-14 & $-1$ & $+20$ & 10.9 \\
2018-05-15 & $-1$ & $-55$ & 10.9 \\
2018-05-16* & $-1$ & $-71$ & 10.8 \\
2018-05-17 & $+1$ & $-26$ & 10.8 \\
2018-05-18* & $+1$ & $-12$ & 11.0 \\
2018-05-19 & $+1$ & $+1$ & 11.0 \\
2018-05-20 & $+1$ & $-40$ & 10.9 \\
2018-05-21* & $+1$ & $+20$ & 10.7 \\
2018-05-22 & $+1$ & $-55$ & 10.9 \\
2018-05-23 & $+1$ & $-71$ & 10.9 \\
2018-05-24* & $-1$ & $-71$ & 10.9 \\
2018-05-25 & 0 & $-26$ & 10.8 \\
2018-05-26 & 0 & $+20$ & 10.9 \\
2018-05-27 & 0 & $-12$ & 11.0 \\
2018-05-28 & 0 & $+1$ & 10.7 \\
2018-05-29* & 0 & $-40$ & 10.7 \\
2018-05-30 & 0 & $-55$ & 10.9 \\
2018-05-31* & 0 & $-71$ & 11.0 \\
2018-06-01 & $+1$ & $-71$ & 11.1 \\
2018-06-03 & 0 & $-26$ & 11.0 \\
2018-06-06* & $-1$ & $-26$ & 9.9 \\
2018-06-12* & $-1$ & $-12$ & 11.0 \\
2018-06-14 & $-1$ & $+1$ & 10.4 \\
2018-06-17* & $-1$ & $-40$ & 11.1 \\
2019-01-27* & 0 & $+1$ & 6.9 \\
2019-01-28* & 0 & $-12$ & 7.2 \\
2019-02-01* & 0 & $+20$ & 7.3 \\
2019-02-02* & 0 & $+1$ & 7.2 \\
2019-02-03 & 0 & $-12$ & 7.1 \\
2019-02-04* & 0 & $+1$ & 7.1 \\
2019-02-05 & 0 & $-12$ & 6.8 \\
2019-02-15 & $+1$ & $+1$ & 9.0 \\
2019-03-16 & $-1$ & $-71$ & 9.0 \\
2019-05-12* & $-1$ & $+1$ & 10.7 \\
2019-05-16* & $-1$ & $-71$ & 10.7 \\
2019-05-18* & $+1$ & $-12$ & 11.0 \\
2019-05-21* & $+1$ & $+20$ & 10.8 \\
2019-05-24* & $-1$ & $-71$ & 10.9 \\
2019-05-29* & 0 & $-40$ & 10.7 \\
2019-05-31 & 0 & $-71$ & 4.5 \\
2019-06-06* & $-1$ & $-26$ & 9.2 \\
2019-06-12* & $-1$ & $-12$ & 10.3 \\
2019-06-17* & $-1$ & $-40$ & 10.1 \\
2020-09-28* & $-1$ & $-71$ & 9.8 \\
2020-09-29 & $-1$ & $-55$ & 9.3 \\
2020-09-30* & $-1$ & $-40$ & 8.2 \\
2020-10-01* & $-1$ & $-26$ & 7.8 \\
2020-10-02 & $-1$ & $-12$ & 7.9 \\
2020-10-03 & $-1$ & $+1$ & 9.8 \\
2020-10-04* & $-1$ & $+20$ & 9.8 \\
2020-10-05 & 0 & $-71$ & 9.8 \\
2020-10-06 & 0 & $-55$ & 9.8 \\
2020-10-07 & 0 & $-40$ & 9.8 \\
2020-10-08 & 0 & $-26$ & 9.8 \\
2020-10-09 & 0 & $-12$ & 9.6 \\
2020-10-10* & 0 & $+1$ & 8.8 \\
2020-10-11 & 0 & $+20$ & 9.8 \\
2020-10-12 & 0 & $-71$ & 8.6 \\
2020-10-13 & 0 & $-55$ & 8.4 \\
2020-10-14 & 0 & $-40$ & 5.6 \\
2020-10-15 & 0 & $-26$ & 9.1 \\
2020-10-16 & 0 & $-12$ & 9.0 \\
2020-10-17* & 0 & $+1$ & 9.8 \\
2020-10-18 & 0 & $+20$ & 9.7 \\
2020-10-19 & $+1$ & $-71$ & 9.5 \\
2020-10-20 & $+1$ & $-55$ & 9.5 \\
2020-10-21 & $+1$ & $-40$ & 9.5 \\
2020-10-22 & $+1$ & $-26$ & 8.2 \\
2020-10-23 & $+1$ & $-12$ & 9.5 \\
2020-10-24* & $+1$ & $+1$ & 9.5 \\
2020-10-25 & $+1$ & $+20$ & 8.0 \\
\hline
& &  \multicolumn{1}{r}{\textbf{Total:}} & 1,056.5\\
\end{supertabular}

\section{Catalogue Column Names}\label{sec:columns}

\input{table.tex}

%GLEAM-X IDR1:C data release:
%\url{https://wiki.mwatelescope.org/display/MP/GLEAM-X\%3A+Internal+Data+Release+1\%3A+Continuum}.

\end{appendix}

\twocolumn

\bibliographystyle{pasa-mnras}
\bibliography{refs}

\end{document}

%% file: table.tex
%\afterpage{
\onecolumn
\topcaption{Column numbers, names, and units for the catalogue.
Source names follow International Astronomical Union naming conventions for co-ordinate-based naming.
Background and RMS measurements were performed by \textsc{BANE} (\Sect~\ref{sec:images});
PSF measurements were peformed using in-house software as described in \Sect~\ref{sec:psf};
the fitted spectral index parameters were derived as described in \Sect~\ref{sec:alpha};
all other measurements were made using \textsc{Aegean}. \textsc{Aegean} incorporates a constrained fitting algorithm.
Shape parameters with an error of $-1$ indicate that the reported value is equal to either the upper or lower fitting constraint.
The columns with the subscript ``wide'' are derived from the 200\,MHz wide-band image.
Subsequently, the subscript indicates the central frequency of the measurement, in MHz.
These sub-band measurements are made using the priorised fitting mode of Aegean, where the position and shape of the source are determined from the wide-band image, and only the flux density is fitted (see \Sect~\ref{sec:detection}).
Note therefore that some columns in the priorised fit do not have error bars, because they are linearly propagated from the wideband image values (e.g. major axis $a$). \label{tab:catalogue}}
\tablefirsthead{\toprule Number & Name & Unit & Description \\ \midrule}
\tablehead{\multicolumn{4}{c}{{\bfseries  Continued from previous page}} \\ \toprule 
Number & Name & Unit & Description \\ \midrule}
\tabletail{ \midrule \multicolumn{4}{c}{{Continued on next page}} \\ \midrule}
\tablelasttail{\bottomrule}
\begin{supertabular}{llll}
1 & Name & hh:mm:ss+dd:mm:ss & International Astronomical Union name  \\ 
2 & background\_wide & Jy\,beam$^{-1}$ & Background in wideband image \\ 
3 & local\_rms\_wide & Jy\,beam$^{-1}$ & Local RMS in wideband image \\ 
4 & ra\_str & hh:mm:ss & Right ascension \\ 
5 & dec\_str & dd:mm:ss & Declination \\ 
6 & RAJ2000 & $\arcdeg$ & Right ascension \\ 
7 & err\_RAJ2000 & $\arcdeg$ & Error on RA \\ 
8 & DEJ2000 & $\arcdeg$ & Declination \\ 
9 & err\_DEJ2000 & $\arcdeg$ & Error on Dec \\ 
10 & peak\_flux\_wide & Jy\,beam$^{-1}$ & Peak flux density in wideband image \\ 
11 & err\_peak\_flux\_wide & Jy\,beam$^{-1}$ & Fitting error on peak flux density in wideband image \\ 
12 & int\_flux\_wide & Jy & Integrated flux density in wideband image \\ 
13 & err\_int\_flux\_wide & Jy & Error on integrated flux density in wideband image \\ 
14 & a\_wide & $\arcsec$ & Major axis of source in wideband image \\ 
15 & err\_a\_wide & $\arcsec$ & Error on major axis of source in wideband image \\ 
16 & b\_wide & $\arcsec$ & Minor axis of source in wideband image \\ 
17 & err\_b\_wide & $\arcsec$ & Error on minor axis of source in wideband image \\ 
18 & pa\_wide & $\arcdeg$ & Postion angle of source in wideband image \\ 
19 & err\_pa\_wide & $\arcdeg$ & Error on position angle of source in wideband image \\ 
20 & residual\_mean\_wide & Jy\,beam$^{-1}$ & Mean of residual after source fitting in wideband image \\ 
21 & residual\_std\_wide & Jy\,beam$^{-1}$ & Standard deviation of residual after source fitting \\ 
22 & err\_abs\_flux\_pct & \% & Percent error in absolute flux scale - all frequencies \\ 
23 & err\_fit\_flux\_pct & \% & Percent error on internal flux scale - all frequencies \\
24 & psf\_a\_wide & $\arcsec$ & Major axis of PSF at location of source in wideband image \\
25 & psf\_b\_wide & $\arcsec$ & Minor axis of PSF at location of source in wideband image \\
26 & psf\_pa\_wide & $\arcdeg$ & Position angle of PSF at location of source in wideband image \\
27 & background\_076 & Jy\,beam$^{-1}$ & Background at 76\,MHz \\
28 & local\_rms\_076 & Jy\,beam$^{-1}$ & Local RMS at 76\,MHz \\
29 & peak\_flux\_076 & Jy\,beam$^{-1}$ & Peak flux density at 76\,MHz \\
30 & err\_peak\_flux\_076 & Jy\,beam$^{-1}$ & Fitting error on peak flux density at 76\,MHz \\
31 & int\_flux\_076 & Jy & Integrated flux density at 76\,MHz \\
32 & err\_int\_flux\_076 & Jy & Fitting error on integrated flux density at 76\,MHz \\
33 & a\_076 & $\arcsec$ & Major axis of source at 76\,MHz \\
34 & b\_076 & $\arcsec$ & Minor axis of source at 76\,MHz \\
35 & pa\_076 & $\arcdeg$ & Position angle of source at 76\,MHz \\
36 & residual\_mean\_076 & Jy\,beam$^{-1}$ & Mean of residual after source fitting at 76\,MHz \\
37 & residual\_std\_076 & Jy\,beam$^{-1}$ & Standard deviation of residual after source fitting at 76\,MHz \\
38 & psf\_a\_076 & $\arcsec$ & Major axis of PSF at location of source at 76\,MHz \\
39 & psf\_b\_076 & $\arcsec$ & Minor axis of PSF at location of source at 76\,MHz \\
40 & psf\_pa\_076 & $\arcdeg$ & Position angle of PSF at location of source at 76\,MHz \\
41 & background\_084 & Jy\,beam$^{-1}$ & Background at 84\,MHz \\
42 & local\_rms\_084 & Jy\,beam$^{-1}$ & Local RMS at 84\,MHz \\
43 & peak\_flux\_084 & Jy\,beam$^{-1}$ & Peak flux density at 84\,MHz \\
44 & err\_peak\_flux\_084 & Jy\,beam$^{-1}$ & Fitting error on peak flux density at 84\,MHz \\
45 & int\_flux\_084 & Jy & Integrated flux density at 84\,MHz \\
46 & err\_int\_flux\_084 & Jy & Fitting error on integrated flux density at 84\,MHz \\
47 & a\_084 & $\arcsec$ & Major axis of source at 84\,MHz \\
48 & b\_084 & $\arcsec$ & Minor axis of source at 84\,MHz \\
49 & pa\_084 & $\arcdeg$ & Position angle of source at 84\,MHz \\
50 & residual\_mean\_084 & Jy\,beam$^{-1}$ & Mean of residual after source fitting at 84\,MHz \\
51 & residual\_std\_084 & Jy\,beam$^{-1}$ & Standard deviation of residual after source fitting at 84\,MHz \\
52 & psf\_a\_084 & $\arcsec$ & Major axis of PSF at location of source at 84\,MHz \\
53 & psf\_b\_084 & $\arcsec$ & Minor axis of PSF at location of source at 84\,MHz \\
54 & psf\_pa\_084 & $\arcdeg$ & Position angle of PSF at location of source at 84\,MHz \\
55 & background\_092 & Jy\,beam$^{-1}$ & Background at 92\,MHz \\
56 & local\_rms\_092 & Jy\,beam$^{-1}$ & Local RMS at 92\,MHz \\
57 & peak\_flux\_092 & Jy\,beam$^{-1}$ & Peak flux density at 92\,MHz \\
58 & err\_peak\_flux\_092 & Jy\,beam$^{-1}$ & Fitting error on peak flux density at 92\,MHz \\
59 & int\_flux\_092 & Jy & Integrated flux density at 92\,MHz \\
60 & err\_int\_flux\_092 & Jy & Fitting error on integrated flux density at 92\,MHz \\
61 & a\_092 & $\arcsec$ & Major axis of source at 92\,MHz \\
62 & b\_092 & $\arcsec$ & Minor axis of source at 92\,MHz \\
63 & pa\_092 & $\arcdeg$ & Position angle of source at 92\,MHz \\
64 & residual\_mean\_092 & Jy\,beam$^{-1}$ & Mean of residual after source fitting at 92\,MHz \\
65 & residual\_std\_092 & Jy\,beam$^{-1}$ & Standard deviation of residual after source fitting at 92\,MHz \\
66 & psf\_a\_092 & $\arcsec$ & Major axis of PSF at location of source at 92\,MHz \\
67 & psf\_b\_092 & $\arcsec$ & Minor axis of PSF at location of source at 92\,MHz \\
68 & psf\_pa\_092 & $\arcdeg$ & Position angle of PSF at location of source at 92\,MHz \\
69 & background\_099 & Jy\,beam$^{-1}$ & Background at 99\,MHz \\
70 & local\_rms\_099 & Jy\,beam$^{-1}$ & Local RMS at 99\,MHz \\
71 & peak\_flux\_099 & Jy\,beam$^{-1}$ & Peak flux density at 99\,MHz \\
72 & err\_peak\_flux\_099 & Jy\,beam$^{-1}$ & Fitting error on peak flux density at 99\,MHz \\
73 & int\_flux\_099 & Jy & Integrated flux density at 99\,MHz \\
74 & err\_int\_flux\_099 & Jy & Fitting error on integrated flux density at 99\,MHz \\
75 & a\_099 & $\arcsec$ & Major axis of source at 99\,MHz \\
76 & b\_099 & $\arcsec$ & Minor axis of source at 99\,MHz \\
77 & pa\_099 & $\arcdeg$ & Position angle of source at 99\,MHz \\
78 & residual\_mean\_099 & Jy\,beam$^{-1}$ & Mean of residual after source fitting at 99\,MHz \\
79 & residual\_std\_099 & Jy\,beam$^{-1}$ & Standard deviation of residual after source fitting at 99\,MHz \\
80 & psf\_a\_099 & $\arcsec$ & Major axis of PSF at location of source at 99\,MHz \\
81 & psf\_b\_099 & $\arcsec$ & Minor axis of PSF at location of source at 99\,MHz \\
82 & psf\_pa\_099 & $\arcdeg$ & Position angle of PSF at location of source at 99\,MHz \\
83 & background\_107 & Jy\,beam$^{-1}$ & Background at 107\,MHz \\
84 & local\_rms\_107 & Jy\,beam$^{-1}$ & Local RMS at 107\,MHz \\
85 & peak\_flux\_107 & Jy\,beam$^{-1}$ & Peak flux density at 107\,MHz \\
86 & err\_peak\_flux\_107 & Jy\,beam$^{-1}$ & Fitting error on peak flux density at 107\,MHz \\
87 & int\_flux\_107 & Jy & Integrated flux density at 107\,MHz \\
88 & err\_int\_flux\_107 & Jy & Fitting error on integrated flux density at 107\,MHz \\
89 & a\_107 & $\arcsec$ & Major axis of source at 107\,MHz \\
90 & b\_107 & $\arcsec$ & Minor axis of source at 107\,MHz \\
91 & pa\_107 & $\arcdeg$ & Position angle of source at 107\,MHz \\
92 & residual\_mean\_107 & Jy\,beam$^{-1}$ & Mean of residual after source fitting at 107\,MHz \\
93 & residual\_std\_107 & Jy\,beam$^{-1}$ & Standard deviation of residual after source fitting at 107\,MHz \\
94 & psf\_a\_107 & $\arcsec$ & Major axis of PSF at location of source at 107\,MHz \\
95 & psf\_b\_107 & $\arcsec$ & Minor axis of PSF at location of source at 107\,MHz \\
96 & psf\_pa\_107 & $\arcdeg$ & Position angle of PSF at location of source at 107\,MHz \\
97 & background\_115 & Jy\,beam$^{-1}$ & Background at 115\,MHz \\
98 & local\_rms\_115 & Jy\,beam$^{-1}$ & Local RMS at 115\,MHz \\
99 & peak\_flux\_115 & Jy\,beam$^{-1}$ & Peak flux density at 115\,MHz \\
100 & err\_peak\_flux\_115 & Jy\,beam$^{-1}$ & Fitting error on peak flux density at 115\,MHz \\
101 & int\_flux\_115 & Jy & Integrated flux density at 115\,MHz \\
102 & err\_int\_flux\_115 & Jy & Fitting error on integrated flux density at 115\,MHz \\
103 & a\_115 & $\arcsec$ & Major axis of source at 115\,MHz \\
104 & b\_115 & $\arcsec$ & Minor axis of source at 115\,MHz \\
105 & pa\_115 & $\arcdeg$ & Position angle of source at 115\,MHz \\
106 & residual\_mean\_115 & Jy\,beam$^{-1}$ & Mean of residual after source fitting at 115\,MHz \\
107 & residual\_std\_115 & Jy\,beam$^{-1}$ & Standard deviation of residual after source fitting at 115\,MHz \\
108 & psf\_a\_115 & $\arcsec$ & Major axis of PSF at location of source at 115\,MHz \\
109 & psf\_b\_115 & $\arcsec$ & Minor axis of PSF at location of source at 115\,MHz \\
110 & psf\_pa\_115 & $\arcdeg$ & Position angle of PSF at location of source at 115\,MHz \\
111 & background\_122 & Jy\,beam$^{-1}$ & Background at 122\,MHz \\
112 & local\_rms\_122 & Jy\,beam$^{-1}$ & Local RMS at 122\,MHz \\
113 & peak\_flux\_122 & Jy\,beam$^{-1}$ & Peak flux density at 122\,MHz \\
114 & err\_peak\_flux\_122 & Jy\,beam$^{-1}$ & Fitting error on peak flux density at 122\,MHz \\
115 & int\_flux\_122 & Jy & Integrated flux density at 122\,MHz \\
116 & err\_int\_flux\_122 & Jy & Fitting error on integrated flux density at 122\,MHz \\
117 & a\_122 & $\arcsec$ & Major axis of source at 122\,MHz \\
118 & b\_122 & $\arcsec$ & Minor axis of source at 122\,MHz \\
119 & pa\_122 & $\arcdeg$ & Position angle of source at 122\,MHz \\
120 & residual\_mean\_122 & Jy\,beam$^{-1}$ & Mean of residual after source fitting at 122\,MHz \\
121 & residual\_std\_122 & Jy\,beam$^{-1}$ & Standard deviation of residual after source fitting at 122\,MHz \\
122 & psf\_a\_122 & $\arcsec$ & Major axis of PSF at location of source at 122\,MHz \\
123 & psf\_b\_122 & $\arcsec$ & Minor axis of PSF at location of source at 122\,MHz \\
124 & psf\_pa\_122 & $\arcdeg$ & Position angle of PSF at location of source at 122\,MHz \\
125 & background\_130 & Jy\,beam$^{-1}$ & Background at 130\,MHz \\
126 & local\_rms\_130 & Jy\,beam$^{-1}$ & Local RMS at 130\,MHz \\
127 & peak\_flux\_130 & Jy\,beam$^{-1}$ & Peak flux density at 130\,MHz \\
128 & err\_peak\_flux\_130 & Jy\,beam$^{-1}$ & Fitting error on peak flux density at 130\,MHz \\
129 & int\_flux\_130 & Jy & Integrated flux density at 130\,MHz \\
130 & err\_int\_flux\_130 & Jy & Fitting error on integrated flux density at 130\,MHz \\
131 & a\_130 & $\arcsec$ & Major axis of source at 130\,MHz \\
132 & b\_130 & $\arcsec$ & Minor axis of source at 130\,MHz \\
133 & pa\_130 & $\arcdeg$ & Position angle of source at 130\,MHz \\
134 & residual\_mean\_130 & Jy\,beam$^{-1}$ & Mean of residual after source fitting at 130\,MHz \\
135 & residual\_std\_130 & Jy\,beam$^{-1}$ & Standard deviation of residual after source fitting at 130\,MHz \\
136 & psf\_a\_130 & $\arcsec$ & Major axis of PSF at location of source at 130\,MHz \\
137 & psf\_b\_130 & $\arcsec$ & Minor axis of PSF at location of source at 130\,MHz \\
138 & psf\_pa\_130 & $\arcdeg$ & Position angle of PSF at location of source at 130\,MHz \\
139 & background\_143 & Jy\,beam$^{-1}$ & Background at 143\,MHz \\
140 & local\_rms\_143 & Jy\,beam$^{-1}$ & Local RMS at 143\,MHz \\
141 & peak\_flux\_143 & Jy\,beam$^{-1}$ & Peak flux density at 143\,MHz \\
142 & err\_peak\_flux\_143 & Jy\,beam$^{-1}$ & Fitting error on peak flux density at 143\,MHz \\
143 & int\_flux\_143 & Jy & Integrated flux density at 143\,MHz \\
144 & err\_int\_flux\_143 & Jy & Fitting error on integrated flux density at 143\,MHz \\
145 & a\_143 & $\arcsec$ & Major axis of source at 143\,MHz \\
146 & b\_143 & $\arcsec$ & Minor axis of source at 143\,MHz \\
147 & pa\_143 & $\arcdeg$ & Position angle of source at 143\,MHz \\
148 & residual\_mean\_143 & Jy\,beam$^{-1}$ & Mean of residual after source fitting at 143\,MHz \\
149 & residual\_std\_143 & Jy\,beam$^{-1}$ & Standard deviation of residual after source fitting at 143\,MHz \\
150 & psf\_a\_143 & $\arcsec$ & Major axis of PSF at location of source at 143\,MHz \\
151 & psf\_b\_143 & $\arcsec$ & Minor axis of PSF at location of source at 143\,MHz \\
152 & psf\_pa\_143 & $\arcdeg$ & Position angle of PSF at location of source at 143\,MHz \\
153 & background\_151 & Jy\,beam$^{-1}$ & Background at 151\,MHz \\
154 & local\_rms\_151 & Jy\,beam$^{-1}$ & Local RMS at 151\,MHz \\
155 & peak\_flux\_151 & Jy\,beam$^{-1}$ & Peak flux density at 151\,MHz \\
156 & err\_peak\_flux\_151 & Jy\,beam$^{-1}$ & Fitting error on peak flux density at 151\,MHz \\
157 & int\_flux\_151 & Jy & Integrated flux density at 151\,MHz \\
158 & err\_int\_flux\_151 & Jy & Fitting error on integrated flux density at 151\,MHz \\
159 & a\_151 & $\arcsec$ & Major axis of source at 151\,MHz \\
160 & b\_151 & $\arcsec$ & Minor axis of source at 151\,MHz \\
161 & pa\_151 & $\arcdeg$ & Position angle of source at 151\,MHz \\
162 & residual\_mean\_151 & Jy\,beam$^{-1}$ & Mean of residual after source fitting at 151\,MHz \\
163 & residual\_std\_151 & Jy\,beam$^{-1}$ & Standard deviation of residual after source fitting at 151\,MHz \\
164 & psf\_a\_151 & $\arcsec$ & Major axis of PSF at location of source at 151\,MHz \\
165 & psf\_b\_151 & $\arcsec$ & Minor axis of PSF at location of source at 151\,MHz \\
166 & psf\_pa\_151 & $\arcdeg$ & Position angle of PSF at location of source at 151\,MHz \\
167 & background\_158 & Jy\,beam$^{-1}$ & Background at 158\,MHz \\
168 & local\_rms\_158 & Jy\,beam$^{-1}$ & Local RMS at 158\,MHz \\
169 & peak\_flux\_158 & Jy\,beam$^{-1}$ & Peak flux density at 158\,MHz \\
170 & err\_peak\_flux\_158 & Jy\,beam$^{-1}$ & Fitting error on peak flux density at 158\,MHz \\
171 & int\_flux\_158 & Jy & Integrated flux density at 158\,MHz \\
172 & err\_int\_flux\_158 & Jy & Fitting error on integrated flux density at 158\,MHz \\
173 & a\_158 & $\arcsec$ & Major axis of source at 158\,MHz \\
174 & b\_158 & $\arcsec$ & Minor axis of source at 158\,MHz \\
175 & pa\_158 & $\arcdeg$ & Position angle of source at 158\,MHz \\
176 & residual\_mean\_158 & Jy\,beam$^{-1}$ & Mean of residual after source fitting at 158\,MHz \\
177 & residual\_std\_158 & Jy\,beam$^{-1}$ & Standard deviation of residual after source fitting at 158\,MHz \\
178 & psf\_a\_158 & $\arcsec$ & Major axis of PSF at location of source at 158\,MHz \\
179 & psf\_b\_158 & $\arcsec$ & Minor axis of PSF at location of source at 158\,MHz \\
180 & psf\_pa\_158 & $\arcdeg$ & Position angle of PSF at location of source at 158\,MHz \\
181 & background\_166 & Jy\,beam$^{-1}$ & Background at 166\,MHz \\
182 & local\_rms\_166 & Jy\,beam$^{-1}$ & Local RMS at 166\,MHz \\
183 & peak\_flux\_166 & Jy\,beam$^{-1}$ & Peak flux density at 166\,MHz \\
184 & err\_peak\_flux\_166 & Jy\,beam$^{-1}$ & Fitting error on peak flux density at 166\,MHz \\
185 & int\_flux\_166 & Jy & Integrated flux density at 166\,MHz \\
186 & err\_int\_flux\_166 & Jy & Fitting error on integrated flux density at 166\,MHz \\
187 & a\_166 & $\arcsec$ & Major axis of source at 166\,MHz \\
188 & b\_166 & $\arcsec$ & Minor axis of source at 166\,MHz \\
189 & pa\_166 & $\arcdeg$ & Position angle of source at 166\,MHz \\
190 & residual\_mean\_166 & Jy\,beam$^{-1}$ & Mean of residual after source fitting at 166\,MHz \\
191 & residual\_std\_166 & Jy\,beam$^{-1}$ & Standard deviation of residual after source fitting at 166\,MHz \\
192 & psf\_a\_166 & $\arcsec$ & Major axis of PSF at location of source at 166\,MHz \\
193 & psf\_b\_166 & $\arcsec$ & Minor axis of PSF at location of source at 166\,MHz \\
194 & psf\_pa\_166 & $\arcdeg$ & Position angle of PSF at location of source at 166\,MHz \\
195 & background\_174 & Jy\,beam$^{-1}$ & Background at 174\,MHz \\
196 & local\_rms\_174 & Jy\,beam$^{-1}$ & Local RMS at 174\,MHz \\
197 & peak\_flux\_174 & Jy\,beam$^{-1}$ & Peak flux density at 174\,MHz \\
198 & err\_peak\_flux\_174 & Jy\,beam$^{-1}$ & Fitting error on peak flux density at 174\,MHz \\
199 & int\_flux\_174 & Jy & Integrated flux density at 174\,MHz \\
200 & err\_int\_flux\_174 & Jy & Fitting error on integrated flux density at 174\,MHz \\
201 & a\_174 & $\arcsec$ & Major axis of source at 174\,MHz \\
202 & b\_174 & $\arcsec$ & Minor axis of source at 174\,MHz \\
203 & pa\_174 & $\arcdeg$ & Position angle of source at 174\,MHz \\
204 & residual\_mean\_174 & Jy\,beam$^{-1}$ & Mean of residual after source fitting at 174\,MHz \\
205 & residual\_std\_174 & Jy\,beam$^{-1}$ & Standard deviation of residual after source fitting at 174\,MHz \\
206 & psf\_a\_174 & $\arcsec$ & Major axis of PSF at location of source at 174\,MHz \\
207 & psf\_b\_174 & $\arcsec$ & Minor axis of PSF at location of source at 174\,MHz \\
208 & psf\_pa\_174 & $\arcdeg$ & Position angle of PSF at location of source at 174\,MHz \\
209 & background\_181 & Jy\,beam$^{-1}$ & Background at 181\,MHz \\
210 & local\_rms\_181 & Jy\,beam$^{-1}$ & Local RMS at 181\,MHz \\
211 & peak\_flux\_181 & Jy\,beam$^{-1}$ & Peak flux density at 181\,MHz \\
212 & err\_peak\_flux\_181 & Jy\,beam$^{-1}$ & Fitting error on peak flux density at 181\,MHz \\
213 & int\_flux\_181 & Jy & Integrated flux density at 181\,MHz \\
214 & err\_int\_flux\_181 & Jy & Fitting error on integrated flux density at 181\,MHz \\
215 & a\_181 & $\arcsec$ & Major axis of source at 181\,MHz \\
216 & b\_181 & $\arcsec$ & Minor axis of source at 181\,MHz \\
217 & pa\_181 & $\arcdeg$ & Position angle of source at 181\,MHz \\
218 & residual\_mean\_181 & Jy\,beam$^{-1}$ & Mean of residual after source fitting at 181\,MHz \\
219 & residual\_std\_181 & Jy\,beam$^{-1}$ & Standard deviation of residual after source fitting at 181\,MHz \\
220 & psf\_a\_181 & $\arcsec$ & Major axis of PSF at location of source at 181\,MHz \\
221 & psf\_b\_181 & $\arcsec$ & Minor axis of PSF at location of source at 181\,MHz \\
222 & psf\_pa\_181 & $\arcdeg$ & Position angle of PSF at location of source at 181\,MHz \\
223 & background\_189 & Jy\,beam$^{-1}$ & Background at 189\,MHz \\
224 & local\_rms\_189 & Jy\,beam$^{-1}$ & Local RMS at 189\,MHz \\
225 & peak\_flux\_189 & Jy\,beam$^{-1}$ & Peak flux density at 189\,MHz \\
226 & err\_peak\_flux\_189 & Jy\,beam$^{-1}$ & Fitting error on peak flux density at 189\,MHz \\
227 & int\_flux\_189 & Jy & Integrated flux density at 189\,MHz \\
228 & err\_int\_flux\_189 & Jy & Fitting error on integrated flux density at 189\,MHz \\
229 & a\_189 & $\arcsec$ & Major axis of source at 189\,MHz \\
230 & b\_189 & $\arcsec$ & Minor axis of source at 189\,MHz \\
231 & pa\_189 & $\arcdeg$ & Position angle of source at 189\,MHz \\
232 & residual\_mean\_189 & Jy\,beam$^{-1}$ & Mean of residual after source fitting at 189\,MHz \\
233 & residual\_std\_189 & Jy\,beam$^{-1}$ & Standard deviation of residual after source fitting at 189\,MHz \\
234 & psf\_a\_189 & $\arcsec$ & Major axis of PSF at location of source at 189\,MHz \\
235 & psf\_b\_189 & $\arcsec$ & Minor axis of PSF at location of source at 189\,MHz \\
236 & psf\_pa\_189 & $\arcdeg$ & Position angle of PSF at location of source at 189\,MHz \\
237 & background\_197 & Jy\,beam$^{-1}$ & Background at 197\,MHz \\
238 & local\_rms\_197 & Jy\,beam$^{-1}$ & Local RMS at 197\,MHz \\
239 & peak\_flux\_197 & Jy\,beam$^{-1}$ & Peak flux density at 197\,MHz \\
240 & err\_peak\_flux\_197 & Jy\,beam$^{-1}$ & Fitting error on peak flux density at 197\,MHz \\
241 & int\_flux\_197 & Jy & Integrated flux density at 197\,MHz \\
242 & err\_int\_flux\_197 & Jy & Fitting error on integrated flux density at 197\,MHz \\
243 & a\_197 & $\arcsec$ & Major axis of source at 197\,MHz \\
244 & b\_197 & $\arcsec$ & Minor axis of source at 197\,MHz \\
245 & pa\_197 & $\arcdeg$ & Position angle of source at 197\,MHz \\
246 & residual\_mean\_197 & Jy\,beam$^{-1}$ & Mean of residual after source fitting at 197\,MHz \\
247 & residual\_std\_197 & Jy\,beam$^{-1}$ & Standard deviation of residual after source fitting at 197\,MHz \\
248 & psf\_a\_197 & $\arcsec$ & Major axis of PSF at location of source at 197\,MHz \\
249 & psf\_b\_197 & $\arcsec$ & Minor axis of PSF at location of source at 197\,MHz \\
250 & psf\_pa\_197 & $\arcdeg$ & Position angle of PSF at location of source at 197\,MHz \\
251 & background\_204 & Jy\,beam$^{-1}$ & Background at 204\,MHz \\
252 & local\_rms\_204 & Jy\,beam$^{-1}$ & Local RMS at 204\,MHz \\
253 & peak\_flux\_204 & Jy\,beam$^{-1}$ & Peak flux density at 204\,MHz \\
254 & err\_peak\_flux\_204 & Jy\,beam$^{-1}$ & Fitting error on peak flux density at 204\,MHz \\
255 & int\_flux\_204 & Jy & Integrated flux density at 204\,MHz \\
256 & err\_int\_flux\_204 & Jy & Fitting error on integrated flux density at 204\,MHz \\
257 & a\_204 & $\arcsec$ & Major axis of source at 204\,MHz \\
258 & b\_204 & $\arcsec$ & Minor axis of source at 204\,MHz \\
259 & pa\_204 & $\arcdeg$ & Position angle of source at 204\,MHz \\
260 & residual\_mean\_204 & Jy\,beam$^{-1}$ & Mean of residual after source fitting at 204\,MHz \\
261 & residual\_std\_204 & Jy\,beam$^{-1}$ & Standard deviation of residual after source fitting at 204\,MHz \\
262 & psf\_a\_204 & $\arcsec$ & Major axis of PSF at location of source at 204\,MHz \\
263 & psf\_b\_204 & $\arcsec$ & Minor axis of PSF at location of source at 204\,MHz \\
264 & psf\_pa\_204 & $\arcdeg$ & Position angle of PSF at location of source at 204\,MHz \\
265 & background\_212 & Jy\,beam$^{-1}$ & Background at 212\,MHz \\
266 & local\_rms\_212 & Jy\,beam$^{-1}$ & Local RMS at 212\,MHz \\
267 & peak\_flux\_212 & Jy\,beam$^{-1}$ & Peak flux density at 212\,MHz \\
268 & err\_peak\_flux\_212 & Jy\,beam$^{-1}$ & Fitting error on peak flux density at 212\,MHz \\
269 & int\_flux\_212 & Jy & Integrated flux density at 212\,MHz \\
270 & err\_int\_flux\_212 & Jy & Fitting error on integrated flux density at 212\,MHz \\
271 & a\_212 & $\arcsec$ & Major axis of source at 212\,MHz \\
272 & b\_212 & $\arcsec$ & Minor axis of source at 212\,MHz \\
273 & pa\_212 & $\arcdeg$ & Position angle of source at 212\,MHz \\
274 & residual\_mean\_212 & Jy\,beam$^{-1}$ & Mean of residual after source fitting at 212\,MHz \\
275 & residual\_std\_212 & Jy\,beam$^{-1}$ & Standard deviation of residual after source fitting at 212\,MHz \\
276 & psf\_a\_212 & $\arcsec$ & Major axis of PSF at location of source at 212\,MHz \\
277 & psf\_b\_212 & $\arcsec$ & Minor axis of PSF at location of source at 212\,MHz \\
278 & psf\_pa\_212 & $\arcdeg$ & Position angle of PSF at location of source at 212\,MHz \\
279 & background\_220 & Jy\,beam$^{-1}$ & Background at 220\,MHz \\
280 & local\_rms\_220 & Jy\,beam$^{-1}$ & Local RMS at 220\,MHz \\
281 & peak\_flux\_220 & Jy\,beam$^{-1}$ & Peak flux density at 220\,MHz \\
282 & err\_peak\_flux\_220 & Jy\,beam$^{-1}$ & Fitting error on peak flux density at 220\,MHz \\
283 & int\_flux\_220 & Jy & Integrated flux density at 220\,MHz \\
284 & err\_int\_flux\_220 & Jy & Fitting error on integrated flux density at 220\,MHz \\
285 & a\_220 & $\arcsec$ & Major axis of source at 220\,MHz \\
286 & b\_220 & $\arcsec$ & Minor axis of source at 220\,MHz \\
287 & pa\_220 & $\arcdeg$ & Position angle of source at 220\,MHz \\
288 & residual\_mean\_220 & Jy\,beam$^{-1}$ & Mean of residual after source fitting at 220\,MHz \\
289 & residual\_std\_220 & Jy\,beam$^{-1}$ & Standard deviation of residual after source fitting at 220\,MHz \\
290 & psf\_a\_220 & $\arcsec$ & Major axis of PSF at location of source at 220\,MHz \\
291 & psf\_b\_220 & $\arcsec$ & Minor axis of PSF at location of source at 220\,MHz \\
292 & psf\_pa\_220 & $\arcdeg$ & Position angle of PSF at location of source at 220\,MHz \\
293 & background\_227 & Jy\,beam$^{-1}$ & Background at 227\,MHz \\
294 & local\_rms\_227 & Jy\,beam$^{-1}$ & Local RMS at 227\,MHz \\
295 & peak\_flux\_227 & Jy\,beam$^{-1}$ & Peak flux density at 227\,MHz \\
296 & err\_peak\_flux\_227 & Jy\,beam$^{-1}$ & Fitting error on peak flux density at 227\,MHz \\
297 & int\_flux\_227 & Jy & Integrated flux density at 227\,MHz \\
298 & err\_int\_flux\_227 & Jy & Fitting error on integrated flux density at 227\,MHz \\
299 & a\_227 & $\arcsec$ & Major axis of source at 227\,MHz \\
300 & b\_227 & $\arcsec$ & Minor axis of source at 227\,MHz \\
301 & pa\_227 & $\arcdeg$ & Position angle of source at 227\,MHz \\
302 & residual\_mean\_227 & Jy\,beam$^{-1}$ & Mean of residual after source fitting at 227\,MHz \\
303 & residual\_std\_227 & Jy\,beam$^{-1}$ & Standard deviation of residual after source fitting at 227\,MHz \\
304 & psf\_a\_227 & $\arcsec$ & Major axis of PSF at location of source at 227\,MHz \\
305 & psf\_b\_227 & $\arcsec$ & Minor axis of PSF at location of source at 227\,MHz \\
306 & psf\_pa\_227 & $\arcdeg$ & Position angle of PSF at location of source at 227\,MHz \\
307 & sp\_int\_flux\_fit\_200 & Jy & Power-law fitted flux density at 200\,MHz \\
308 & err\_sp\_int\_flux\_fit\_200 & Jy & Error on power-law fitted flux density at 200\,MHz \\
309 & sp\_alpha & -- & Fitted spectral index assuming a power-law SED \\
310 & err\_sp\_alpha & -- & Error on power-law fitted spectral index \\
311 & sp\_reduced\_chi2 & -- & Reduced $\chi^2$ statistic for power-law SED fit \\
312 & csp\_int\_flux\_fit\_200 & Jy & Curved SED fitted flux density at 200\,MHz \\
313 & err\_csp\_int\_flux\_fit\_200 & Jy & Error on curved SED fitted flux density at 200\,MHz \\
314 & csp\_alpha & -- & Fitted spectral index assuming a curved SED \\
315 & err\_csp\_alpha & -- & Error on curved SED fitted spectral index \\
316 & csp\_beta & -- & Fitted curvature index for curved SED fit \\
317 & err\_csp\_beta & -- & Error on curvature index for curved SED fit \\
318 & csp\_reduced\_chi2 & -- & Reduced $\chi^2$ statistic for curved SED fit \\
\end{supertabular}